\documentclass[review,authoryear]{elsarticle} 

\journal{ }
\graphicspath{{./Figures/}}

\usepackage{amsmath,amssymb}
\usepackage{bm}
\usepackage{gensymb}
\usepackage[colorlinks]{hyperref}
\usepackage{subcaption}
\usepackage{booktabs}

\newcommand{ \h }{ \ensuremath{ \bar{h} } }
\newcommand{ \T }{ \ensuremath{ \overline{T} } }
\newcommand{\uu}{\ensuremath{\bar\upsilon}}
\newcommand{\dd}{\ensuremath{\delta}}
\newcommand{\ee}{\ensuremath{\epsilon}}
\newcommand{\cc}{\ensuremath{\bar{\chi}}}
\DeclareMathOperator{\sgn}{sgn}

\newcommand{\thm}{\ensuremath{_{\mathrm{th}}}}

\newcommand{\tpose}{\ensuremath{^{\top}}}
\newcommand{\vect}[1]{\ensuremath{\bm{{\rm #1}}}}

\begin{document}

\begin{frontmatter}

\title{The machine is the material: Structures that mimic one-dimensional thermoelastic materials}

\author[ACCIS]{Matthew P. O'Donnell\corref{cor1}}
\author[ACCIS]{Jonathan P. Stacey} 
\author[SoM]{Isaac V. Chenchiah} 
\author[ACCIS]{Alberto~Pirrera} 

\address[ACCIS]{Bristol Composites Institute (ACCIS), University of Bristol, Queen's Building, Bristol UK, BS8 1TR}
\address[SoM]{School of Mathematics, University of Bristol, University Walk, Bristol UK, BS8 1TW}

\cortext[cor1]{Corresponding Author: Matt.ODonnell@bristol.ac.uk}

\begin{abstract}
Materials that behave like machines, e.g.\ functional materials that are able to change shape in response to external stimuli \citep{Bhattacharya_James:2005}, often do so by exploiting phase transitions. Shape memory materials and the tail sheath of Bacteriophage T4 are two well-known examples. For the resulting machine to be effective, the material needs to have desirable and tunable properties. Developing such materials has proven to be an endeavour which requires considerable expertise in materials science, engineering and mathematics \citep{ZHANG20094332}.

Here, we reverse this approach by instead designing a machine that acts as a material. Our methodology is independent of characteristic length, allowing us to design behaviour from the architected material through to the macroscopic scale. Specifically, we present thermally-actuated structures whose effective continuum behaviour is that of one-dimensional thermoelastic materials. We show that these structures may possess a range of behaviours, such as shape memory, zero or negative thermal expansivity. Moreover, the amplitude 

of the behaviour, e.g.\ length change at critical temperature or magnitude of thermal expansivity, can exceed what is attainable through conventional materials. Seemingly incompatible features, such as low barriers to transformation and stiffness across high elongations, can be combined; the designer can independently control the critical heating and cooling temperatures and eliminate hysteresis, if desired; changes in length can be either continuous or discontinuous; and shape memory can be combined with negative thermal expansivity.
\end{abstract}


\end{frontmatter}

\section{Introduction}

Advanced structures often need internal mechanisms to achieve adaptability. Instead, structures made of transformative materials can adapt to the environment without requiring complex actuation mechanisms. The potential advantages offered by transformative materials have lead to significant efforts to tailor them for adaptivity. Shape memory alloys (SMAs) are a well-known example~\citep{Bhattacharya_James:2005}. SMAs and other multi-physics-based actuators have been extensively investigated for industrial use~\citep{Sigmund:2001,Frecker:2003,Hartl_Lagoudas:2007}. In general, there have been significant efforts to tune the thermal properties of materials and structures for bespoke adaptivity~\citep{Sigmund_Torquato:1996,Miller2009, Tao_et_al:2015,Boatti_et_al:2017}. Nevertheless, despite the many successes of materials-based approaches, there remain limitations on the performance achievable, e.g.\ small strains, temperature hysteresis and low stiffness.

Herein, instead of attempting to tailor material properties to obtain desired macroscopic responses, we embed functionality at the structural component level; see, for instance,~\citet{Bertoldi:2017}. By coupling anisotropic stiffness properties with nonlinear geometric deformations we can achieve desired adaptive macroscopic behaviours, whilst also significantly expanding the range of possible responses beyond what is achievable through material tailoring alone.

Including nonlinear geometric effects to obtain functional structural behaviours~\citep{Wang_et_al:2013,  Fraternali_et_al:2015,GROH2018394} reflects a 
design paradigm which is seen across various length scales from microstructure tailoring~\citep{Xia:2017} to large morphing/deployable systems~\citep{Pirrera_et_al:2010, Chen_et_al:2012, Dewalque_et_al:2018}. Of specific relevance to 
this paper are composite lattice structures that have been studied as part of hierarchical design concepts~\citep{ODonnell_et_al:2016}. 
 
The present paper is based on the behaviour of the composite helical lattice structure proposed by~\citet{Pirrera_et_al:2013} and inspired by the virus Bacteriophage T4. This lattice utilises geometry, anisotropy, and pre-stress as a source of nonlinearity and the ensuing functionality. Its elastic response was shown to be robust, repeatable, and tailorable, making it a candidate for the development of either passively or actively actuated architected materials.

In this work, we consider temperature-dependent behaviour, which provides an additional means for actuation. Thermal stresses act to modify the pre-stress of the lattice, modifying its energetic landscape. The desired temperature-dependent responses may be achieved through a careful balance of the stiffness of the strips that the lattice is comprised of, their thermal response, and their mechanical pre-stress. Specifically, we present thermally-actuated one-dimensional structures that can:
\begin{itemize}
\item behave as SMAs at the structural length scale. 
\item be tailored to create negative thermal expansivity or other unusual thermal expansion profiles.
\end{itemize}

Similar thermal responses arising from nonlinear geometric reconfigurations are observed across length scales. At the material level, mechanisms in crystalline lattices which lead to negative thermal expansivities are analogous to the geometric effects of the helical structure proposed here \citep{Barrera_et_al:2005}. At the structural level, by tailoring the void space of hybrid lattice structures, \citet{Jefferson_et_al:2009} achieve a variety of tuneable thermal responses that differ significantly from the expansion behaviour of the constituent parts. However, it is often difficult to tune a structure's or a material's thermal response whilst also maintaining a desirable stiffness profile. In our approach, the potential to obtain better trade-offs is increased, because geometry, mechanical pre-stress and material anisotropy can be designed synergistically to give the lattice specific effective stiffness properties. 

We now proceed to outline the mathematical framework underpinning the behaviour of the thermally-adaptive lattice structures presented herein. Examples of the range of responses that can be obtained are discussed along with the results obtained from a proof-of-concept demonstrator. It is noted that thermal actuation is just one possible route to adaptivity. Alternatives such as piezoelectricity~\citep{Tawfik_et_al:2011} and magnetoelectricity~\citep{Palneedi_et_al:2016} could readily be integrated into the present model.

\section{A thermoelastic helical lattice structure}
\label{sec:sec2}

\citet{Pirrera_et_al:2013} developed an analytical energy-based formulation to describe the behaviour of helical lattices comprised of anisotropic strips. Here, in order to embed thermal responsivity, we exploit the material's extension-curvature couplings and amplify thermally-induced deformations kinematically through the structural arrangement. A simple system such as a  bi-morph, which is a strip composed of two materials with different coefficients of thermal expansions (CTEs), is sufficient to show the occurrence of such thermally-induced deformations. More generally, anisotropic materials offer increased tailorability and so are included in our formulation.

\subsection{Lattice Geometry}

\begin{figure*}
  \centering
  \begin{subfigure}[t]{.32\textwidth}
  \centering
  \includegraphics[width=\linewidth]{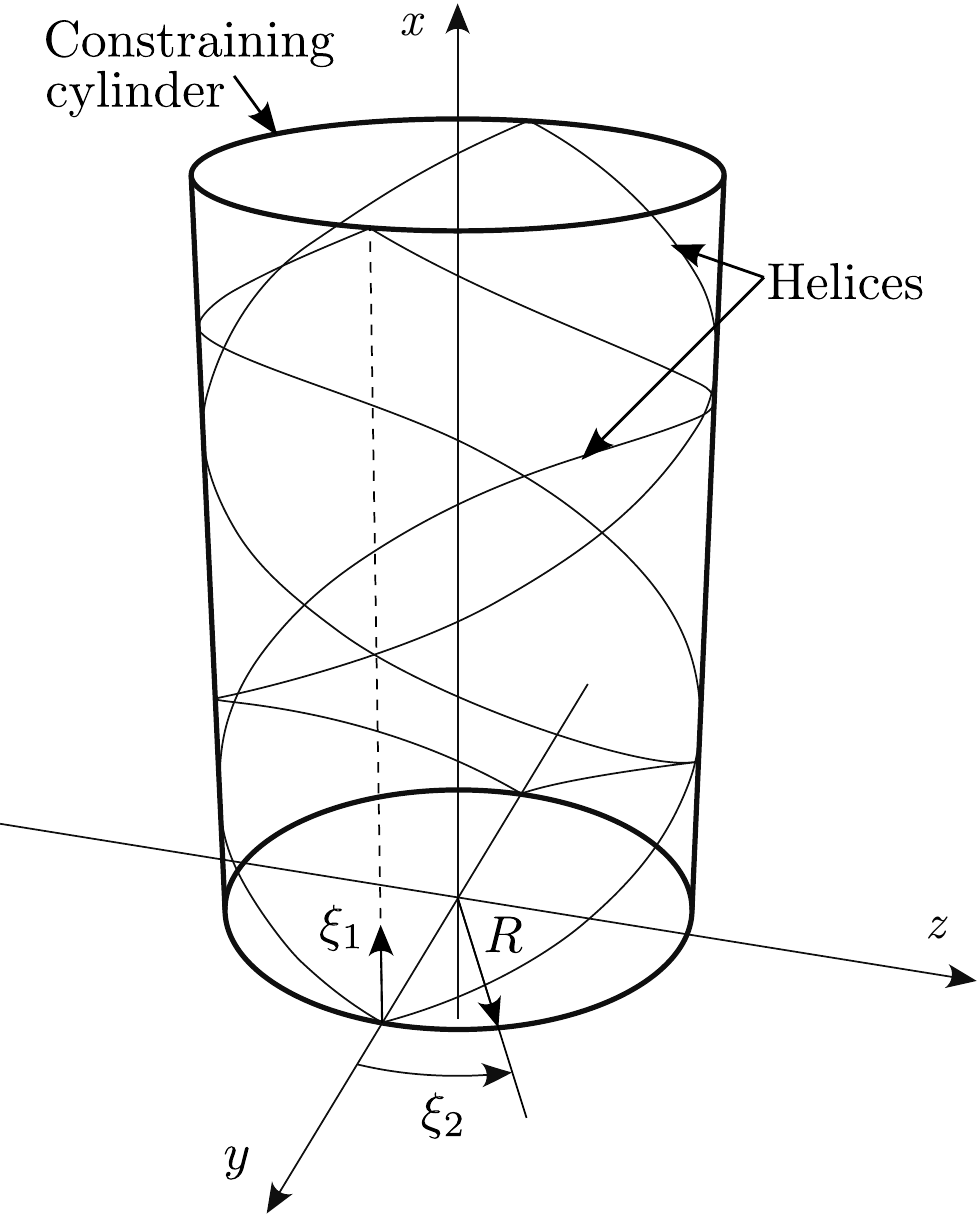}
  \caption{A cylindrical lattice.}
  \end{subfigure} 
  \hfill
  \begin{subfigure}[t]{.32\textwidth}
  \centering
  \includegraphics[width=\linewidth]{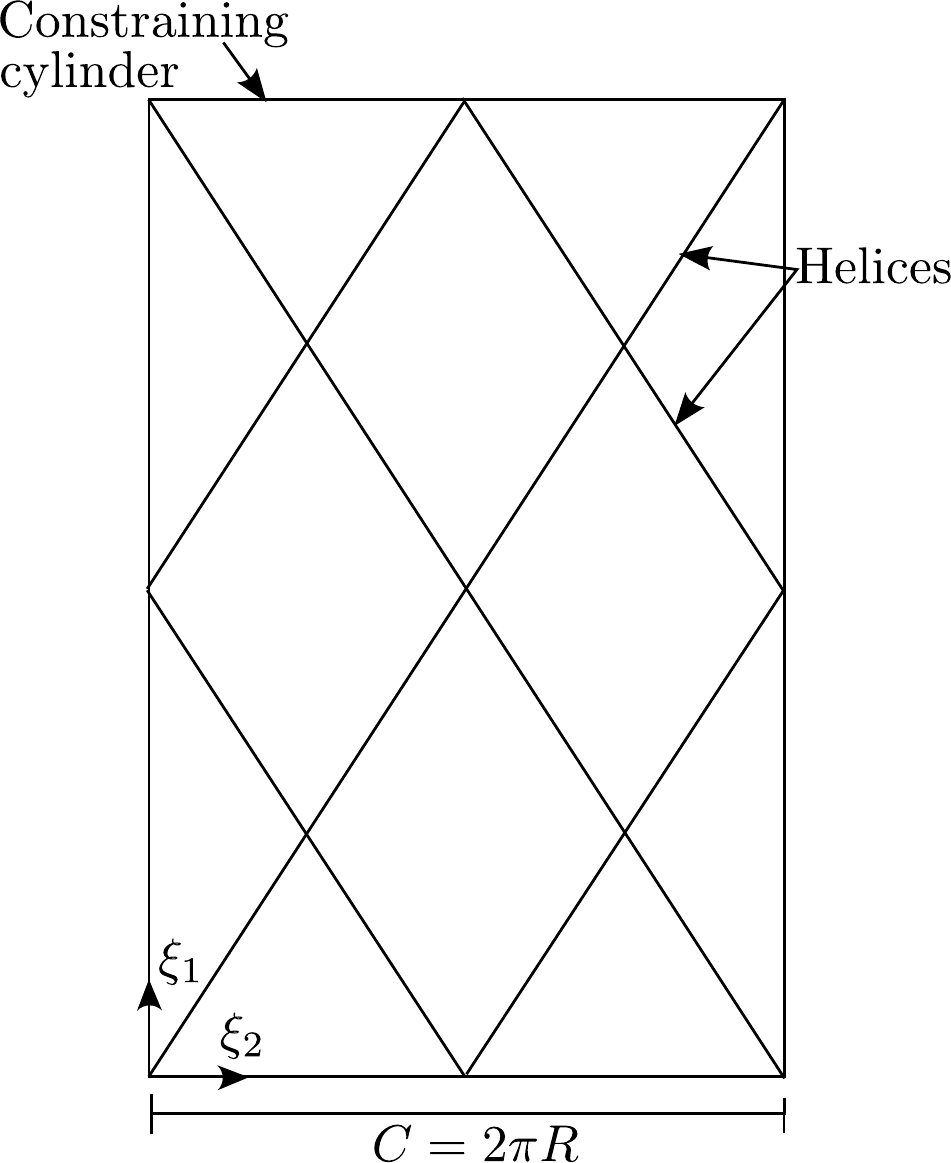}
      \caption{Planar projection of lattice.}
  \end{subfigure}  
  \hfill
  \begin{subfigure}[t]{.32\textwidth}
  \centering
  \includegraphics[width=\linewidth]{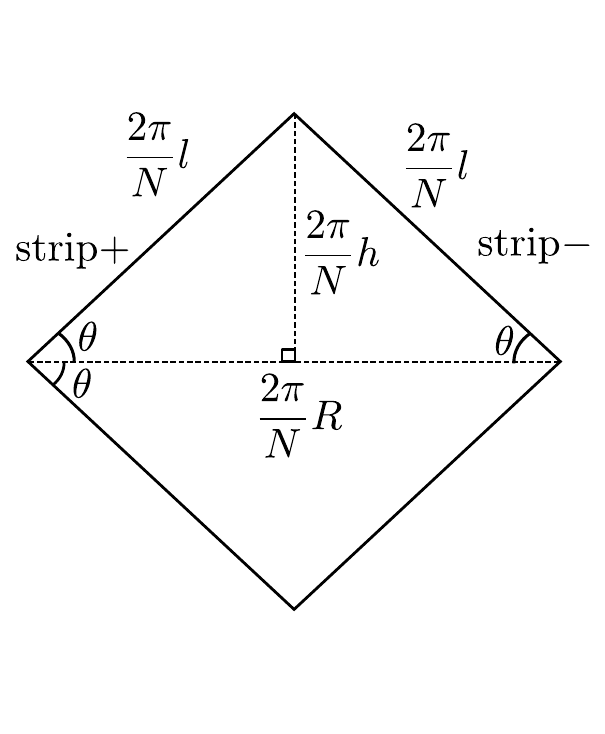}
      \caption{Rhomboidal unit cell.}
  \end{subfigure}
      \caption{Sketch of a helical lattice comprised of $N=2$ strips of each handednes on the surface of a constraining cylinder of radius $R$, and its planar representation.}
      \label{fig:HelixSketch}
\end{figure*}

The lattice is represented schematically in Figure~\ref{fig:HelixSketch} and comprises of $N$ helices of each handedness. Herein, in contrast to \citet{Pirrera_et_al:2013},  we assume a rhomboidal unit cell. This restriction prevents the lattice from twisting upon elongation or contraction. Consequently, the lattice extension, $h$, characterises the system uniquely~\citep{ODonnell_et_al:2016}. The horizontal diagonal of the unit cell is $\frac{2\pi}{N} R$, where $R$ is the radius of the helices, and the vertical diagonal is $\frac{4\pi}{N} h$. It is convenient to set the side length to be $\frac{2\pi}{N} l$. 

Referring to Figure~\ref{fig:HelixSketch}, all right-handed lattice strips have an identical stiffness and mechanical pre-curvature. So do the left-handed strips, but right- and left-handed strips could differ from each other. Henceforth, right- and left-handed strips are denoted by subscripts $\bullet_+$ and $\bullet_-$, respectively. 

Making the same kinematic assumptions as~\cite{Pirrera_et_al:2013}, we now turn to the energy-based formulation of lattice behaviour including thermal effects. These assumptions are:
\begin{itemize}
\item all helices remain on a single cylinder (whose radius and height can change);
\item the non-thermal change in length of the helical strips is small and can be neglected;
\item the helices are hinged where they intersect allowing only scissoring motion. 
\end{itemize}

\subsection{Total Potential Energy}
As in~\citet{Pirrera_et_al:2013}, we represent generic anisotropic material properties recurring to Classical Laminate Theory (CLT) for its convenient notation.
CLT also provides a robust framework to capture the thermoelastic response of the structure under consideration. 

Let us consider a generic laminate, with the coordinate system $(x,y,z)$ chosen such that $z$ is the thickness direction. As is conventional, cf.\ e.g.\ \cite{Mansfield:2005}, we assume that layers of a laminated composite are perfectly bonded to each other, and that the stiffness properties may be condensed to the mid-plane. It follows that, for the $k^\text{th}$ layer, which is assumed to be orthotropic, the stress-strain relation expressed in the laminate coordinates is
\begin{equation} \label{eq:stress}
\boldsymbol{\sigma}^{(k)}
 = \bar{\boldsymbol{Q}}^{(k)} \left( \boldsymbol{\epsilon} - \Delta T \boldsymbol{\alpha}^{(k)} \right),
\end{equation}
where, as usual, $\boldsymbol{\epsilon} = \boldsymbol{\varepsilon} + z \boldsymbol{\kappa}$ and the temperature change, $\Delta T$, is $T - T_{\text{ref}}$, $T$ being the current temperature and $T_{\text{ref}}$ a reference temperature corresponding to the stress-free state. For the $k^\text{th}$ layer, $\boldsymbol{\alpha}^{(k)} = [ \alpha_{xx}, \alpha_{yy}, 2 \alpha_{xy} ]\tpose$ is the vector coefficient of thermal expansion and $\bar{\boldsymbol{Q}}^{(k)}$ is the transformed stiffness matrix.

Integrating~\eqref{eq:stress} through the thickness, we obtain resultant forces and moments per unit width, $\mathbf{N}$ and $\mathbf{M}$, which are related to strains, $\boldsymbol{\varepsilon}$, curvatures, $\boldsymbol{\kappa}$, and $\Delta T$, through
\begin{equation} \label{eq:therm_clt}
\begin{bmatrix} \mathbf{N} \\ \mathbf{M} \end{bmatrix}
 = \begin{bmatrix} \mathbf{A} & \mathbf{B} \\ \mathbf{B} & \mathbf{D} \end{bmatrix}
       \begin{bmatrix} \boldsymbol{\varepsilon} \\ \boldsymbol{\kappa} \end{bmatrix}
       - \Delta T  \begin{bmatrix} \mathbf{N}\thm \\ \mathbf{M}\thm \end{bmatrix},
\end{equation}
where $\mathbf{A}$, $\mathbf{B}$ and $\mathbf{D}$ are stiffness matrices, and $\mathbf{N}\thm$ and $\mathbf{M}\thm$ are thermal stress resultants. 

The strain energy of the $k^\text{th}$ layer, per unit surface area, i.e.\ before integration through the strips' width and length, is
\begin{equation} \label{eq:layer_energy}
U^{(k)}
 = \frac{1}{2} \int_{z_k}^{z_{k+1}} \left( \boldsymbol{\epsilon} - \Delta T \boldsymbol{\alpha}^{(k)} \right)\tpose \bar{\boldsymbol{Q}}^{(k)} \left( \boldsymbol{\epsilon} - \Delta T \boldsymbol{\alpha}^{(k)} \right) \, \mathrm{d}z,
\end{equation}
where $z_k$ and $z_{k+1}$ are the positions of the interfaces between layers. Integrating~\eqref{eq:layer_energy} through thickness, we obtain the total strain energy per unit surface area to be
\begin{equation} \label{eq:energy}
U 
 = \frac{1}{2} \begin{bmatrix} \boldsymbol{\varepsilon} \\ \boldsymbol{\kappa} \end{bmatrix}\tpose 
     \begin{bmatrix} \mathbf{A} & \mathbf{B} \\ \mathbf{B} & \mathbf{D} \end{bmatrix}
      \begin{bmatrix} \boldsymbol{\varepsilon} \\ \boldsymbol{\kappa} \end{bmatrix}
     - \Delta T \begin{bmatrix} \mathbf{N}\thm \\ \mathbf{M}\thm \end{bmatrix}\tpose
        \begin{bmatrix} \boldsymbol{\varepsilon} \\ \boldsymbol{\kappa} \end{bmatrix}
     + \frac{U\thm}{2} (\Delta T)^2,
\end{equation}
where
\begin{equation*}
U\thm
 = \sum_k \int_{z_k}^{z_{k+1}}  {\boldsymbol{\alpha}^{(k)}}\tpose \bar{\boldsymbol{Q}}^{(k)} \boldsymbol{\alpha}^{(k)} \, \mathrm{d}z.
\end{equation*}

As in~\citet{Pirrera_et_al:2013}, we wish to write the energy in terms of $\boldsymbol{\kappa}$, $\mathbf{N}$ and $\mathbf{N}\thm$. To do this we use the partially-inverted form of~\eqref{eq:therm_clt}, which is
\begin{equation} \label{eq:local5}
\begin{bmatrix}
\boldsymbol{\varepsilon}  \\
\mathbf{M} +  \Delta T \mathbf{M}\thm\\
\end{bmatrix}
 =
\begin{bmatrix}
\mathbf{a} & \mathbf{b} \\
-\mathbf{b}\tpose & \mathbf{d}
\end{bmatrix} 
\begin{bmatrix}
\mathbf{N}  +  \Delta T \mathbf{N}\thm\\
\boldsymbol{\kappa}
\end{bmatrix},
\end{equation}
with
\begin{subequations}
\begin{align}\label{eq:abddefs}
\mathbf{a} &=  \mathbf{A}^{-1},  \\
\mathbf{b} &= -\mathbf{A}^{-1}\mathbf{B}, \\
\mathbf{d} &=  \mathbf{D}-\mathbf{B}\mathbf{A}^{-1}\mathbf{B}.
\end{align}
\end{subequations}
Substituting~\eqref{eq:local5} in~\eqref{eq:energy}, we obtain
\begin{equation*}
U
 = \frac{1}{2} \boldsymbol{\kappa}\tpose \mathbf{d} \boldsymbol{\kappa}
     - \left( \mathbf{N}\thm\tpose \mathbf{b} + \mathbf{M}\thm\tpose \right) \mathbf{d} \boldsymbol{\kappa} \Delta T 
     + \frac{1}{2} \left( ( \mathbf{N} - \mathbf{N}\thm )\tpose \mathbf{a} ( \mathbf{N} + \mathbf{N}\thm ) + U\thm \right) (\Delta T)^2.
\end{equation*}
Assuming, as in~\citet{Pirrera_et_al:2013}, zero in-plane stress resultants, i.e.\ $\mathbf{N}= \mathbf{0}$, the strain energy simplifies to
\begin{equation*}
U
 = \frac{1}{2} \boldsymbol{\kappa}\tpose \mathbf{d} \boldsymbol{\kappa}
     - \left( \mathbf{b}\tpose \mathbf{N}\thm + \mathbf{M}\thm \right)\tpose \mathbf{d} \boldsymbol{\kappa} \Delta T 
     + \frac{1}{2} \left( U\thm - \mathbf{N}\thm\tpose \mathbf{a} \mathbf{N}\thm \right) (\Delta T)^2.
\end{equation*}
Further, it is convenient to set
\begin{subequations}
\begin{align}
\vect{\chi}
 &= \mathbf{b}\tpose \, \mathbf{N}\thm + \mathbf{M}\thm, \\
\tau
 &= U\thm - \mathbf{N}\thm\tpose \mathbf{a} \mathbf{N}\thm,\label{eq:tau}
\end{align}
\end{subequations}
which leads to
\begin{equation}\label{eq:energyperunit}
U
 = \frac{1}{2} \boldsymbol{\kappa}\tpose \mathbf{d} \boldsymbol{\kappa}
     - \vect{\chi}\tpose \mathbf{d} \boldsymbol{\kappa} \Delta T 
     + \frac{\tau}{2} (\Delta T)^2.
\end{equation}
We now present the energy formulation in terms of the lattice's representative unit cell.

\subsection{Energy of a unit cell}

With reference to Figure~\ref{fig:HelixSketch}, the strain energy of the representative unit cell, $\Pi$, follows from~\eqref{eq:energyperunit}, whence,
\begin{equation} \label{eq:energydef}
\frac{N}{2\pi} \Pi
 = \begin{aligned}[t]
     &\frac{l}{2} \left(w_{+} \boldsymbol{\kappa}_{+}^{\top}\mathbf{d}_{+}\boldsymbol{\kappa}_{+}  + w_{-} \boldsymbol{\kappa}_{-}^{\top}\mathbf{d}_{-}\boldsymbol{\kappa}_{-} \right) \\
     &\quad - l \left( w_+ \vect{\chi}_+\tpose \mathbf{d}_+ \boldsymbol{\kappa}_{+} + w_- \vect{\chi}_-\tpose \mathbf{d}_- \boldsymbol{\kappa}_{-} \right) \Delta T \\
     &\qquad + \frac{l}{2}\left(w_{+}\tau_{+} + w_{-}\tau_{-} \right) (\Delta T)^2, 
    \end{aligned} 
\end{equation}
where  $\frac{2\pi}{N} l$ is the side length of the unit cell and $w_{\pm}$ the strips' widths. 

Following~\citet{Pirrera_et_al:2013}, we consider only axial and twist components of curvature. These can be written to account for a reference mechanical stress-free curvature, e.g.\ a manufacturing tooling curvature, as
\begin{align}
\boldsymbol{\kappa}_{\pm}
 &= \begin{bmatrix} \kappa_{x\pm}(h) \\ \kappa_{xy\pm}(h) \end{bmatrix}
       - \begin{bmatrix} \upsilon_{x\pm} \\ \upsilon_{xy\pm} \end{bmatrix},
\end{align}
where, again from~\citet{Pirrera_et_al:2013},
\begin{equation}
\begin{bmatrix} \kappa_{x\pm}(h) \\ \kappa_{xy\pm}(h) \end{bmatrix}
 =  \frac{1}{R}
\begin{bmatrix}
\cos^2\theta \\
\pm \sin \theta \cos \theta
\end{bmatrix}
=  
\frac{1}{2l^2} 
\begin{bmatrix}
\sqrt{l^2 - h^2} \\
\pm h
\end{bmatrix},
\end{equation}  
and where $R$ is the radius of the constraining cylinder and $\theta$ the helical angle as illustrated in Figure~\ref{fig:HelixSketch}.

The bending stiffness, accounting only for axial and twist components of curvature, reduces to
\begin{equation*}
\mathbf{d}_{\pm}
 = \begin{bmatrix} d_{11\pm} & d_{16\pm} \\ d_{16\pm} & d_{66\pm} \end{bmatrix},
\end{equation*}
and $\vect{\chi}$ reduces to
\begin{equation*}
\vect{\chi}
 = \begin{bmatrix} \chi_x \\ \chi_{xy} \end{bmatrix}.
\end{equation*}

We now proceed to express the unit cell energy in terms of non-dimensional parameters.

\section{Normalised energy landscape}
\label{sec:sec3}

We now begin our analysis of the helical lattice, which we will henceforth view as an one-dimensional object whose state is specified by its non-dimensional length and non-dimensional temperature,
\begin{subequations} \label{eq:non-dimensional}
\begin{align}
\h
 &= \frac{h}{l} \in [0,1], \\
\T
 &= \frac{\Delta T}{T_{\text{ref}}},
\end{align} 
\end{subequations}
respectively. We remark that $l$ might depend on temperature through the in-plane thermal expansion effects captured by $\tau$, however such temperature-induced changes are subsumed into $\h$ (see also Section~\ref{sec:thermal_invariance} below) .

\subsection{The energy density}

From~\eqref{eq:energydef} and~\eqref{eq:non-dimensional}, the non-dimensionalised energy density, i.e.\ energy per unit cell, is given by
\begin{align}
\bar{\Pi}
 &= \frac{4 N l}{\pi w_+ d_{11+} }\Pi, \\
 &= \begin{aligned}[t]
     &{a}_{00} + {a}_{01} \T + {a}_{02} \T^2 + {a}_{10} \h + {a}_{11} \h\T + {a}_{20} \h^2 \\
     &\qquad + \sqrt{1-\h^2} \left( {b}_{10} + {b}_{11} \T  + {b}_{20} \h \right)
    \end{aligned} \label{eq:NDenergy}
\end{align}
with 
\begin{subequations} \label{eq:coefficients}
\begin{align}
{a}_{00}
&= \begin{aligned}[t]  
    &1 + \uu_{x+}^2 + 2\ee_{+}\uu_{x+}\uu_{xy+} + \dd_{+}\uu_{xy+}^2 \\
    &+ \varphi \left(1 + \uu_{x-}^2 + 2\ee_{-}\uu_{x-}\uu_{xy-} + \dd_{-}\uu_{xy-}^2 \right),
   \end{aligned} \\
{a}_{01}
&= \begin{aligned}[t]
    2 T_{\text{ref}} & \left( \left( \cc_{x+} \uu_{x+} +\ee_{+} \cc_{x+} \uu_{xy+} +   \ee_{+} \cc_{xy+} \uu_{x+} + \dd_{+} \cc_{xy+} \uu_{xy+} \right) \right. \\
    &\left. + \varphi \left( \cc_{x-} \uu_{x-} + \ee_{-} \cc_{x-} \uu_{xy-} + \ee_{-} \cc_{xy-} \uu_{x-} + \dd_{-} \cc_{xy-} \uu_{xy-} \right) \right),
   \end{aligned} \\
{a}_{02}
&= 4 \frac{l^2 T_{\text{ref}}^2\tau_{+}}{d_{11+}}
\left( 1 + \frac{ w_{-} \tau_{-} }{ w_{+} \tau_{+} } \right), \label{eq:in-plane-thermal} \\
{a}_{10}
&= 2 \left( \varphi \left( \ee_{-} \uu_{x-} + \dd_{-} \uu_{xy-} \right) - \left( \ee_{+} \uu_{x+} + \dd_{+} \uu_{xy+} \right) \right), \\
{a}_{11} 
&=  2 T_{\text{ref}} \left(     
\varphi \left( \ee_{-} \cc_{x-} + \dd_{-} \cc_{xy-} \right)
-\left( \ee_{+} \cc_{x+} - \dd_{+} \cc_{xy+} \right) \right), \\
{a}_{20}
&=  \left( \dd_{+} -1 \right) + \varphi \left( \dd_{-} - 1 \right), \\
{b}_{10} 
&= - 2 \left( \left( \uu_{x+} + \ee_{+} \uu_{xy+} \right) + \varphi \left( \uu_{x-} +  \ee_{-} \uu_{xy-} \right) \right), \\
{b}_{11}
& = -2 T_{\text{ref}} \left( \left( \cc_{x+} + \ee_{+} \cc_{xy+} \right) + \varphi \left(
\cc_{x-} + \ee_{-} \cc_{xy-} \right) \right), \\
{b}_{20} & = 2 \left( \ee_{+} - \varphi \ee_{-} \right), \label{eq:in-plane}
\end{align}
\end{subequations}
and
\begin{subequations}
\begin{align}
\cc_{\pm}
 &= 2 l \boldsymbol{\chi}_{\pm}, \\
\uu_{\pm}
 &= 2 l \boldsymbol{\upsilon}_{\pm}, \\
\varphi
 &= \frac{w_{-}d_{11-}}{w_{+}d_{11+}} > 0, \\ 
\bar{\mathbf{d}}_{\pm}
 &= \frac{\mathbf{d}}{d_{11\pm}}
 = \begin{bmatrix} 1 & \ee_{\pm} \\ \ee_{\pm} & \dd_{\pm} \end{bmatrix}. \label{eq:local1}
\end{align}
\end{subequations}
For the stiffness matrix in~\eqref{eq:local1} to be positive definite, we need $\dd_{\pm}>0$ and $-\sqrt{\dd_{\pm}}<\ee_{\pm}<\sqrt{\dd_{\pm}}$.

\subsection{Equilibria of the energy density}
\label{sec:equilibria}

To identify system equilibria we compute the derivative of the energy density with respect to extension. From~\eqref{eq:NDenergy},
\begin{equation} \label{eq:diff1a}
\frac{\partial\bar{\Pi}}{\partial \h}
 = {a}_{10} + {a}_{11} \T + 2 {a}_{20} \h
  + \frac{1}{\sqrt{1-\h^2}} \left( {b}_{20} - {b}_{10} \h - {b}_{11} \h \T - 2 {b}_{20} \h^2 \right).
\end{equation}

\paragraph{Inner equilibria}
The location and stability of \emph{inner} equilibria, i.e.\ equilibria for $\h \in (0,1)$, may be obtained by setting $\frac{\partial\bar{\Pi}}{\partial \h} = 0$ in~\eqref{eq:diff1a}.
Temperature-invariant equilibria are obtained when ${a}_{11} = {b}_{11} = 0$. Otherwise, the equilibria generically form a quartic curve in $(\h,\T) \in (0,1) \times \mathbb{R}$ (generically cubic if ${a}_{20} = {b}_{20}$). We denote stable equilibria at temperature $\T$ (if any) by $\h_0(\T)$.

\paragraph{Boundary equilibria}
The stability of the fully-coiled state, $\h = 0$, is governed by
\begin{subequations} \label{eq:diff1b}
\begin{equation}\label{eq:stability0}
\left.\frac{\partial \bar{\Pi}}{\partial \h}\right|_{\h=0}
 = {b}_{20} + {a}_{10} + {a}_{11}\T;
\end{equation}
and that of the fully-extended state, $\h = 1$, by
\begin{equation}\label{eq:stability1}
\begin{cases}
\sgn \left. \frac{\partial \bar{\Pi}}{\partial \h} \right|_{\h \to 1}
 = - \sgn ({b}_{10} + {b}_{20} + {b}_{11} \T)
  & \text{if} \quad {b}_{10} + {b}_{20} + {b}_{11} \T \neq 0, \\
\left.\frac{\partial \bar{\Pi}}{\partial \h}\right|_{\h = 1}
 = {a}_{10} + 2 {a}_{20} + {a}_{11} \T
  &\text{if} \quad {b}_{10} + {b}_{20} + {b}_{11} \T = 0.
\end{cases}
\end{equation}
\end{subequations} 

As above, temperature-invariant equilibria are obtained when ${a}_{11} = {b}_{11} = 0$. On the other hand, when ${a}_{11} \neq 0$ and ${b}_{11} \neq 0$, from the linear dependence on $\T$ in~\eqref{eq:stability0} and~\eqref{eq:stability1}, we conclude that the stability of each boundary changes (i.e.\ from stable to unstable or vice-versa) precisely once if the temperature is varied monotonically. This occurs at the temperatures
\begin{subequations} \label{eq:actuation_temperature}
\begin{align}
\T_0
 &= - \frac{{a}_{10} + {b}_{20}}{{a}_{11}}, \label{eq:actuation_temperature0} \\
\T_1
 &= - \frac{{b}_{10} + {b}_{20} }{{b}_{11}}, \label{eq:actuation_temperature1}
\end{align}
\end{subequations}
for $\h = 0$ and $\h = 1$, respectively. Whether the change is from stable to unstable or vice-versa depends on sign of ${a}_{11}$ and ${b}_{11}$ (and, of course, the direction of the temperature change). Note also that~\eqref{eq:stability0} vanishes 
at $\T = \T_0$, but the stability of $\h = 1$ at $\T = \T_1$ depends on the sign of ${a}_{10} + 2 {a}_{20} + {a}_{11} \T_1$ (cf.,\ \eqref{eq:stability1} and~\eqref{eq:actuation_temperature1}).

In summary, changes in stability of the boundary solutions  occur precisely at critical temperatures $\T=\T_0$ and $\T=\T_1$, for $\h=0$ and $\h=1$, respectively. From this observation we can infer that, during a temperature sweep, once a critical temperature is passed, the stability of the boundary does not change further.  Boundary solutions will remain stable (or unstable) until the direction of the temperature sweep is reversed and the critical temperatures are passed again. This behaviour is beneficial from a thermal actuation standpoint, because it ensures that the system remains in one state once actuation has occurred.

\subsection{Extensional stiffness}

The effective stiffness of the system is the second derivative of the energy with respect to the extension. From~\eqref{eq:diff1a},
\begin{equation} \label{eq:diff2}
\frac{\partial^2\bar{\Pi}}{\partial \h^2}
 = 2 {a}_{20} + \frac{1}{\left(1-\h^2\right)^{3/2}} \left( - {b}_{10} - {b}_{11} \T - 3 {b}_{20} \h + 2 {b}_{20} \h^3 \right).
\end{equation}

We remark that it is possible, using~\eqref{eq:diff1a} and~\eqref{eq:diff2},  to tune the lattice to possess a zero stiffness state at a specified temperature. This feature can be useful for actuation by means of input parameters other than temperature, as it minimises the energy required for shape changes induced, for example, by mechanical or piezoelectric stimuli.

At the boundaries, we obtain
\begin{equation*}
\left. \frac{\partial^2\bar{\Pi}}{\partial \h^2} \right|_{\h = 0}
 = 2 {a}_{20} - {b}_{10} - {b}_{11} \T.
\end{equation*}
and
\begin{equation*}
\begin{cases}
\sgn \left. \frac{\partial^2\bar{\Pi}}{\partial \h^2} \right|_{\h \to 1}
 = - \sgn ({b}_{10} + {b}_{20} + {b}_{11} \T)
  & \text{if} \quad {b}_{10} + {b}_{20} +  {b}_{11} \T \neq 0, \\
\left. \frac{\partial^2\bar{\Pi}}{\partial \h^2} \right|_{\h = 1}
 = 2 {a}_{20}
  & \text{if} \quad {b}_{10} + {b}_{20} + {b}_{11} \T = 0.
\end{cases}
\end{equation*}

Similar to the equilibria above, we see that there is no temperature dependence if ${b}_{11} = 0$. Otherwise, the sign of the stiffness at $\h = 0$ changes at temperature $\T_0'$ given by 
\begin{equation}
\T_0'
 = \frac{ 2 {a}_{20} - {b}_{10} }{ {b}_{11} }.
\end{equation}
However, the sign of the stiffness at $\h = 1$ changes at $\T_1$
defined in ~\eqref{eq:actuation_temperature1}. Thus, unstable equilibria are possible at $\h = 0$, but not at $\h = 1$ unless, $\T = \T_1$ and ${a}_{20} < 0$.

\subsection{Coefficient of Thermal Expansion}
\label{sec:CTE}

The non-dimensional coefficient of thermal expansion (CTE) for the system may be defined as
\begin{equation} \label{eq:CTE}
 \frac{\partial \h_{0}}{\partial \T}, 
\end{equation}
where $\h_0 \in (0,1)$ is a solution of~\eqref{eq:diff1a}. An explicit expression for the CTE can be obtained from~\eqref{eq:diff1a} but, for brevity, we prefer to present numerical examples in Section~\ref{sec:responses} below.

In order to avoid apparently infinite thermal expansions when $\h_0$ approaches zero, ~\eqref{eq:CTE} omits the reciprocal of initial length $\tfrac{1}{\h_0}$ and, instead, a nominal value of one is used corresponding to the total non-dimensional length of the lattice strip. (However, when discussing experimental results the usual definition of CTE is applied.) Equation \eqref{eq:CTE} is non-zero only when either ${a}_{11} \neq 0$ or ${b}_{11} \neq 0$. Moreover the CTE can be zero at $\h_0 = 0$ and $\h_0 = 1$ due to the stability conditions of the boundary (cf.\ last paragraph of Section~\ref{sec:equilibria}).

We remark that the CTE referred to here neglects changes in length of the structure due to the effect of temperature on $l$, whose effect is characterised by $\tau$ defined in~\eqref{eq:tau}. (See also Section~\ref{sec:thermal_invariance} below.)

\section{Thermoelastic Behaviour}
\label{sec:responses}

The thermoelastic response of the lattice may be tailored to fall into two categories, snapping actuation and smooth actuation.
In the first instance, the lattice undergoes a large extension/contraction upon heating/cooling to critical temperatures. Such sudden transitions occur when an equilibrium looses stability, forcing the system to reconfigure by snapping across the energy landscape onto a new stable configuration. Alternatively, extension and contraction can be smooth
transitions. In these instances, stable equilibria change position gradually, maintaining their stability as the temperature varies. 

Either kind of response may be desirable. Snapping actuation, for example, can replicate the typical behaviour of SMAs. However, as we shall see below, the lattice's critical temperatures, direction of actuation, effective stiffness and presence of hysteresis between cooling and heating actuation points, can all be simultaneously tuned far beyond that currently possible for SMAs.

Next, we present several examples of both kinds of responses. 
\subsection{Snap-actuation with hysteresis}
\label{sec:Forward}

As our first example, we illustrate snap actuation from fully-coiled to fully-extended configurations. The transition is realised upon heating to a critical temperature, $\T_0$, whilst snap-back to the coiled state happens upon cooling to a critical temperature, $\T_1$, which should not be higher than $\T_0$; recalling that $\T_0$ and $\T_1$ are defined in~\eqref{eq:actuation_temperature}. Such an actuator mimics an SMA's behaviour.

For this behaviour we require:
\begin{itemize}

\item The fully-coiled configuration, $\h = 0$, to be stable if and only if $\T < \T_0$, and the fully-extended configuration, $\h = 1$, to be stable if and only if $\T > \T_1$. From~\eqref{eq:diff1b} and~\eqref{eq:actuation_temperature}, this implies,
\begin{subequations} \label{eq:snapping}
\begin{align}
{a}_{11} 
 &< 0, \\
{b}_{11} 
 &> 0,
\end{align}
\end{subequations}
respectively.

\item The temperature above which the fully-coiled configuration loses stability, $\T_0$, to be greater than the temperature above which the fully-extended state gains stability, $\T_1$:
\begin{equation} \label{eq:local2}
\T_0
 > \T_1.
\end{equation}

\item In addition, for the equilibrium to shift from $\h = 0$ to $\h = 1$ as the temperature is increased to $\T_0$, we require $$\frac{\partial\bar{\Pi}(T_0,h)}{\partial \h} < 0 \quad \forall \, \h \in (0,1),$$ which, from~\eqref{eq:diff1a} and~\eqref{eq:actuation_temperature0}, can be cast as 
\begin{subequations} \label{eq:actuation1}
\begin{multline}
-{b}_{20} + 2 {a}_{20} \h
 + \frac{1}{\sqrt{1-\h^2}}
     \left( {b}_{20} - {b}_{10} \h + {b}_{11} \left( \frac{ {a}_{10} + {b}_{20} }{ {a}_{11} } \right) \h - 2 {b}_{20} \h^2 \right)
 < 0, \\
 \forall  \, \h \in (0,1).
\end{multline}

\item Similarly, for the equilibrium to shift from $\h = 1$ to $\h = 0$ as the temperature is decreased to $\T_1$, we require $$\frac{\partial\bar{\Pi}(T_1,h)}{\partial \h} > 0 \quad \forall \, \h \in (0,1),$$ or equivalently, using~\eqref{eq:diff1a} and~\eqref{eq:actuation_temperature1}, that
\begin{multline}
{a}_{10} - {a}_{11}\left(\frac{{b}_{20} + {b}_{10} }{{b}_{11}} \right)
+2{a}_{20}\h + \frac{{b}_{20}}{\sqrt{1-\h^2}}
\left(1 + \h - 2\h^2\right)
 > 0, \\
 \forall \, \h \in (0,1).
\end{multline}
\end{subequations}

\end{itemize}

In summary, equations~\eqref{eq:snapping}, \eqref{eq:local2} and~\eqref{eq:actuation1} are the conditions under which the system exhibits hysteretic snap actuation.

If, as a special case which simplifies~\eqref{eq:actuation1} by removing the dependancy on $\h$, we assume that
\begin{subequations} \label{eq:actuation2}
\begin{align}
{a}_{20}
 &= 0, \\
{b}_{20}
 &= 0,
\end{align} 
\end{subequations}
then~\eqref{eq:actuation1} is equivalent to
\begin{subequations} \label{eq:local3}
\begin{align}
- {b}_{10} + {b}_{11} \frac{ {a}_{10}  }{ {a}_{11} }
 &< 0, \\
{a}_{10} - {a}_{11} \frac{ {b}_{10} }{ {b}_{11} }
 &> 0.
\end{align}
\end{subequations}

In other words, when~\eqref{eq:actuation2} holds, equations~\eqref{eq:snapping}, \eqref{eq:local2} and~\eqref{eq:actuation1} can be replaced by equations~\eqref{eq:snapping}, \eqref{eq:local2} and \eqref{eq:local3}. However, using~\eqref{eq:actuation_temperature}, it is easy to verify that~\eqref{eq:snapping} and~\eqref{eq:local2} imply~\eqref{eq:local3}. Thus, equations~\eqref{eq:snapping}, \eqref{eq:local2} and~\eqref{eq:actuation1} can be replaced by equations~\eqref{eq:snapping}, \eqref{eq:local2} and \eqref{eq:actuation2}.

\paragraph{Numerical example}
We now present a numerical example of actuation with hysteresis which satisfies~\eqref{eq:actuation2}. With no loss of generality, we choose the simplest parameters that lead to~\eqref{eq:actuation2}, namely,
\begin{subequations} \label{eq:parameters1}
\begin{align}
\delta_\pm
 &= 1, \\
\epsilon_\pm
 &=0.
\end{align}
In addition it is convenient to set
\begin{equation}
\varphi
 = 1.
\end{equation}
\end{subequations}
Hence, from~\eqref{eq:coefficients} and~\eqref{eq:actuation_temperature},
\begin{align*}
{a}_{11}
 &= -2T_{\text{ref}} \left( \bar{\chi}_{xy+} - \bar{\chi}_{xy-} \right), \\
{b}_{11}
 &= -2 T_{\text{ref}} \left( \bar{\chi}_{x+} + \bar{\chi}_{x-} \right), \\
\T_0
 &=  - \frac{ {a}_{10} }{ {a}_{11} }
   = -\frac{1}{ T_{\text{ref}} }  \frac{ \bar{\upsilon}_{xy+} - \bar{\upsilon}_{xy-} }{ \bar{\chi}_{xy+} - \bar{\chi}_{xy-} }, \\
\T_1
 &= - \frac{ {b}_{10} }{ {b}_{11} }
   = - \frac{1}{ T_{\text{ref}} } \frac{ \bar{\upsilon}_{x+} + \bar{\upsilon}_{x-} }{ \bar{\chi}_{x+} + \bar{\chi}_{x-} }. 
\end{align*}
To satisfy~\eqref{eq:snapping}, \eqref{eq:local2} and restricting $\T_0$ and $\T_1$ to be negative, we need,
\begin{subequations} \label{eq:local4}
\begin{align}
-\left(\bar{\chi}_{x+} + \bar{\chi}_{x-}\right)
 &> 0, \label{eq:local4a} \\
\bar{\chi}_{xy+} - \bar{\chi}_{xy-}
 &> 0, \label{eq:local4b} \\
\bar{\upsilon}_{xy+} - \bar{\upsilon}_{xy-}
 &> 0, \\
-\left(\bar{\upsilon}_{x+} + \bar{\upsilon}_{x-}\right)
 &> 0, \\
\frac{ \bar{\upsilon}_{xy+} - \bar{\upsilon}_{xy-} }{ \bar{\chi}_{xy+} - \bar{\chi}_{xy-} }
 &> - \frac{ \bar{\upsilon}_{x+} + \bar{\upsilon}_{x-} }{ \bar{\chi}_{x+} + \bar{\chi}_{x-} }.
\end{align}
\end{subequations}
Note, as a consequence of~\eqref{eq:local4}, that the lattice cannot possess chiral 
symmetry. 

For this and subsequent numerical examples, we choose design parameters as listed in Table~\ref{tab:properties}; the table also lists the tuned values of $\T_0$ and $\T_1$. We also choose
\begin{subequations} \label{eq:parameters2}
\begin{equation}
T_{\text{ref}}
 = 453 K.
\end{equation}
These values are illustrative, rather than restrictive, and are selected to demonstrate system responses that appear robustly in parameter space while facilitating analysis by simplifying~\eqref{eq:coefficients}. In addition, the choice of parameters ensures realistic material properties and feasible structural designs.

Inspecting the energy density~\eqref{eq:NDenergy}, we observe from~\eqref{eq:in-plane-thermal} that ${a}_{02}$ is the only term associated with in-plane thermal expansion parameters, $\tau_\pm$. The corresponding term is uncoupled from $\h$. As a consequence, it does not affect the position of equilibria, i.e.\  ${a}_{02}$ is not present in equation \eqref{eq:diff1a}.  As a consequence it also does not affect the qualitative response of the system or the effective coefficient of thermal expansion, see~\eqref{eq:CTE} below. The value of ${a}_{02}$ can therefore be specified arbitrarily and, for convenience, we pick
\begin{equation}
{a}_{02} 
 = 1.
\end{equation}
\end{subequations}

\begin{table}[]
\centering
\footnotesize
\begin{tabular}{@{}rrrrrrrrrrrr@{}}
\toprule
& & & &
\multicolumn{8}{|c|}{All curvature parameters scaled by $\times10^4$}  \\
Sec.\ & 
Fig.\ &
$\T_0$ &
$\T_1$ &
$\bar{\chi}_{x}^{+}$ &
$\bar{\chi}_{x}^{-}$ &
$\bar{\nu}_{x}^{+}$ &
$\bar{\nu}_{x}^{-}$ &
$\bar{\chi}_{xy}^{+}$ &
$\bar{\chi}_{xy}^{-}$
& $\bar{\nu}_{xy}^{+}$
& $\bar{\nu}_{xy}^{-}$
\\ \midrule
\ref{sec:Forward} & 
\ref{fig:Forward} & 
$-0.2$ & 
$-0.3   $ & 
$-1$ & 
$-1$ & 
$\bar{T}_1 T_{\text{ref}}$ & 
$\bar{T}_1 T_{\text{ref}}$ & 
$2$ &  
$1$ &  
$-2\bar{T}_0T_{\text{ref}}$ & 
$-\bar{T}_0T_{\text{ref}}$ 
\\
%
\ref{sec:NoH} & 
\ref{fig:NoH} & 
$-0.2$ & 
$-0.2$ & 
$-1$ & 
$-1$ & 
$\bar{T}_1 T_{\text{ref}}$ & 
$\bar{T}_1 T_{\text{ref}}$ & 
$2$ &  
$1$ &  
$-2\bar{T}_0T_{\text{ref}}$ & 
$-\bar{T}_0T_{\text{ref}}$ 
\\
%
\ref{sec:NoH} & 
\ref{fig:NoHR} & 
$-0.2$ & 
$-0.2$ & 
$1$ & 
$1$ & 
$-\bar{T}_1 T_{\text{ref}}$ & 
$-\bar{T}_1 T_{\text{ref}}$ & 
$-2$ &  
$-1$ &  
$2\bar{T}_0T_{\text{ref}}$ & 
$\bar{T}_0T_{\text{ref}}$ 
\\
%
\ref{sec:Smooth} & 
\ref{fig:Smooth} & 
$-0.35$ & 
$0$ & 
$-1$ & 
$-1$ & 
$\bar{T}_1 T_{\text{ref}}$ & 
$\bar{T}_1 T_{\text{ref}}$ & 
$2$ &  
$1$ &  
$-2\bar{T}_0T_{\text{ref}}$ & 
$-\bar{T}_0T_{\text{ref}}$ 
\\
\ref{sec:Smooth} & 
\ref{fig:SmoothR} & 
$0.0$ & 
$-0.35$ & 
$1$ & 
$1$ & 
$-\bar{T}_1 T_{\text{ref}}$ & 
$-\bar{T}_1 T_{\text{ref}}$ & 
$-2$ &  
$-1$ &  
$2\bar{T}_0T_{\text{ref}}$ & 
$\bar{T}_0T_{\text{ref}}$ 
\\
\ref{sec:Smooth} & 
\ref{fig:Critical} & 
$-0.1$ & 
$-0.1$ & 
$-1$ & 
$-1$ & 
$\bar{T}_0 T_{\text{ref}}$ & 
$\bar{T}_0 T_{\text{ref}}$ & 
$-4$ &  
$-2$ &  
$4\bar{T}_0T_{\text{ref}}$ & 
$2\bar{T}_0T_{\text{ref}}$ 
\\
 \bottomrule
\end{tabular}
\caption{System parameters for the examples presented.}
\label{tab:properties}
\end{table}

\begin{figure*}
\centering
    \begin{subfigure}[b]{0.48\textwidth}
        \includegraphics[width=\textwidth]{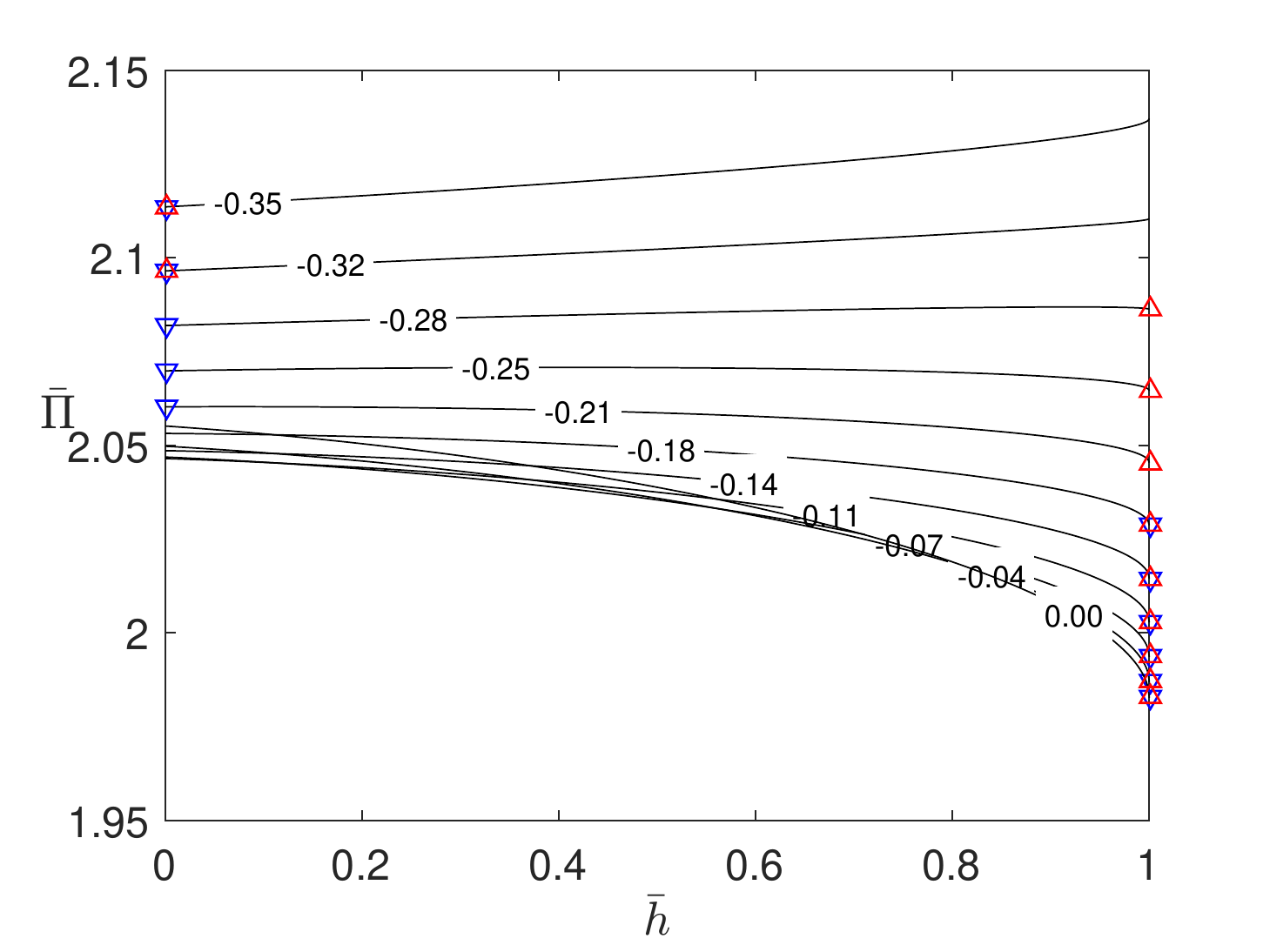}
        \caption{{\footnotesize Energy profiles  for varying $\bar{T}$ with stable equilibria at various temperatures.}}
        \label{sfig:ForwardEnergy}
    \end{subfigure}
    ~
    \begin{subfigure}[b]{0.48\textwidth}
        \includegraphics[width=\textwidth]{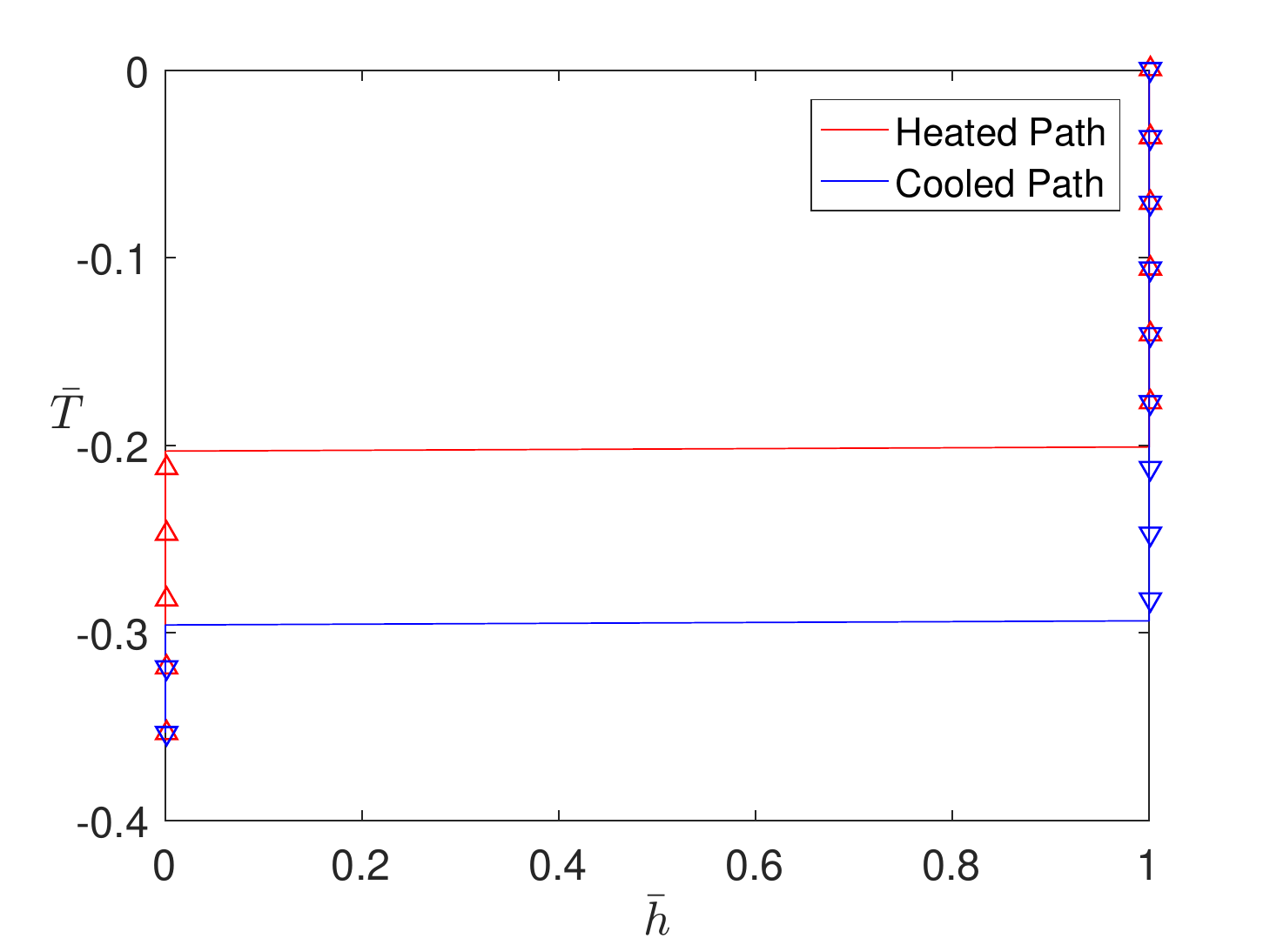}
        \caption{{\footnotesize Location of equilibrium during heating and cooling.}}
        \label{sfig:ForwardPos}
    \end{subfigure}   
    \caption{Actuation from fully coiled to fully expanded at $\T = -0.2$ and recoil at $\T = -0.3$.}
    \label{fig:Forward}
\end{figure*}

Returning to the example in this section, it is easy to verify that the selected parameters satisfy~\eqref{eq:local4}. 
The resulting snap-actuation response is shown in Figure~\ref{fig:Forward}. Figure~\ref{sfig:ForwardPos} demonstrates that, for increasing temperature, $\h = 0$ is stable until $\T = \T_0 = -0.2$. On the other hand, upon cooling, $\h = 1$ looses stability at $\T = \T_1 = -0.3$. Consequently the structure demonstrates hysteresis, being bi-stable in the range $[\T_1, \T_0] = [-0.3,-0.2]$ and monostable elsewhere.

\subsection{Snap-actuation with no hysteresis}
\label{sec:NoH}

\begin{figure*}
\centering
    \begin{subfigure}[t]{0.48\textwidth}
        \includegraphics[width=\textwidth]{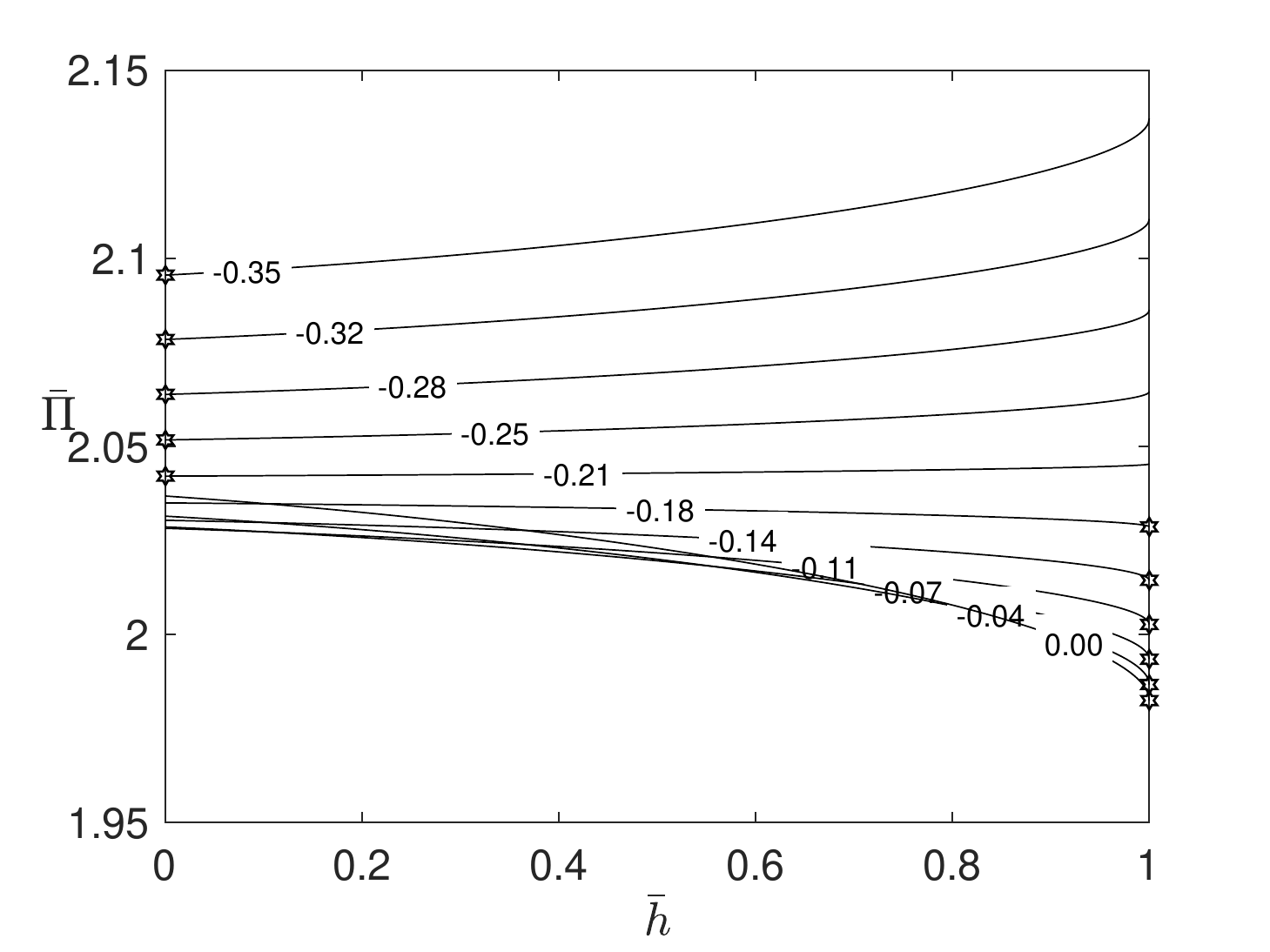}
        \caption{{\footnotesize Energy profiles for varying $\bar{T}$ with stable equilibria at various temperatures.}}
        \label{sfig:NoHEnergy}
    \end{subfigure}
    ~
    \begin{subfigure}[t]{0.48\textwidth}
        \includegraphics[width=\textwidth]{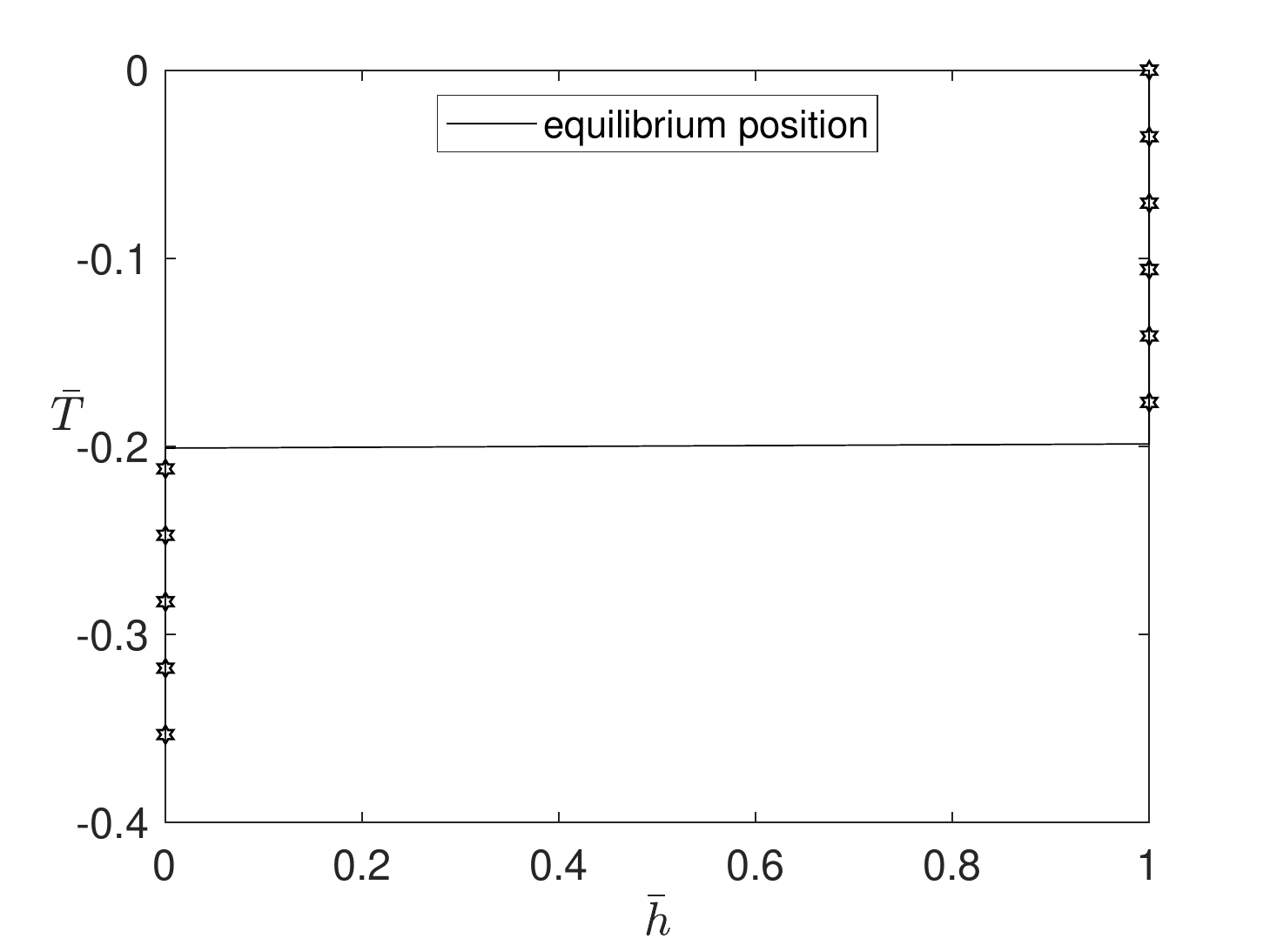}
        \caption{{\footnotesize Location of equilibrium during heating and cooling. The paths coincide.}}
        \label{sfig:NoHPos}
    \end{subfigure}   
    \caption{Expansion actuation without hysteresis at $\T = -0.2$.}
    \label{fig:NoH}
\end{figure*}

The removal of thermal hysteresis for SMAs requires subtle fine-tuning of the atomic lattice \citep{Cui2006,ZHANG20094332,Song2013}. In contrast, the system presented herein can be readily tuned to remove hysteresis by setting $\T_0 = \T_1$. Such a lattice would require
\begin{gather*}
\T_0 
 = -\frac{{a}_{10}}{{a}_{11}} = - \frac{{b}_{10}}{{b}_{11}} = \T_1, \\
{a}_{20} 
 = {b}_{20} = 0, \\
{a}_{11}
 <0, \\
{b}_{11}
 > 0. 
\end{gather*}
Figure~\ref{fig:NoH} presents an example of such behaviour with the parameters given as before and in Table~\ref{tab:properties}.

\begin{figure*}
\centering
    \begin{subfigure}[t]{0.48\textwidth}
        \includegraphics[width=\textwidth]{newNoHE}
        \caption{{\footnotesize Energy profiles  for varying $\bar{T}$ with stable equilibria at various temperatures.}}
        \label{sfig:NoHREnergy}
    \end{subfigure}
    ~
    \begin{subfigure}[t]{0.48\textwidth}
        \includegraphics[width=\textwidth]{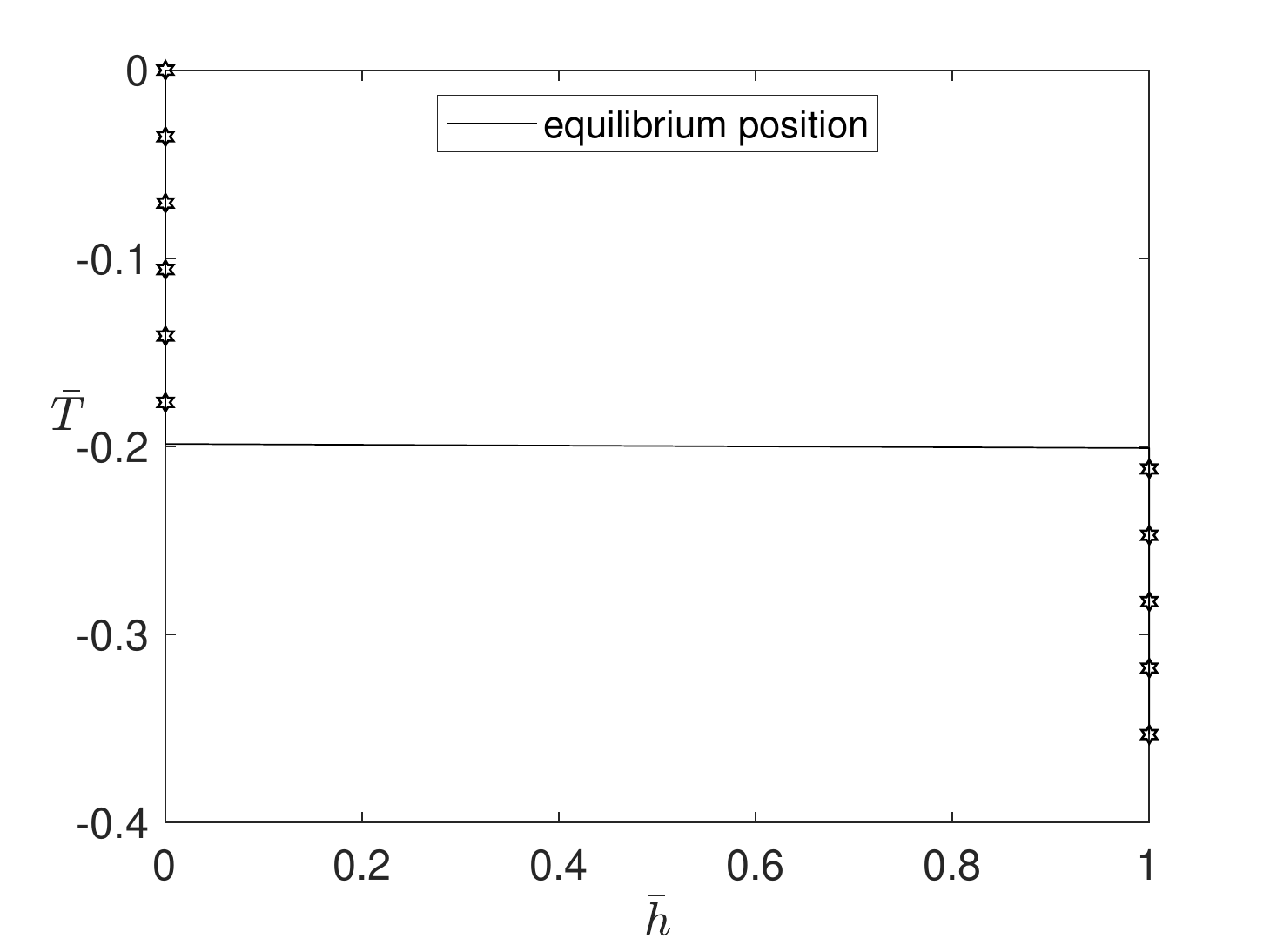}
        \caption{\footnotesize Location of stable equilibrium during heating and cooling. The paths coincide.}
        \label{sfig:NoHRPos}
    \end{subfigure}   
    \caption{{\footnotesize Contraction actuation without hysterisis at $\T = -0.2$.}}
    \label{fig:NoHR}
\end{figure*}

Alternatively, the direction of actuation may be reversed by choosing $a_{11}>0$ and $b_{11}<0$, as explained in Section~\ref{sec:equilibria}. We can therefore devise a system with no hysteresis that snaps from fully-extended to fully-coiled. This is illustrated in Figure~\ref{fig:NoHR}.

To conclude this example, we note that a lattice with no hysteresis has zero stiffness at the transition temperature, as illustrated by the flat energy landscape in Figures~\ref{sfig:NoHEnergy} and~\ref{sfig:NoHREnergy}.

\subsection{Smooth actuation}
\label{sec:Smooth}

\begin{figure*}
\centering
    \begin{subfigure}[b]{0.48\textwidth}
        \includegraphics[width=\textwidth]{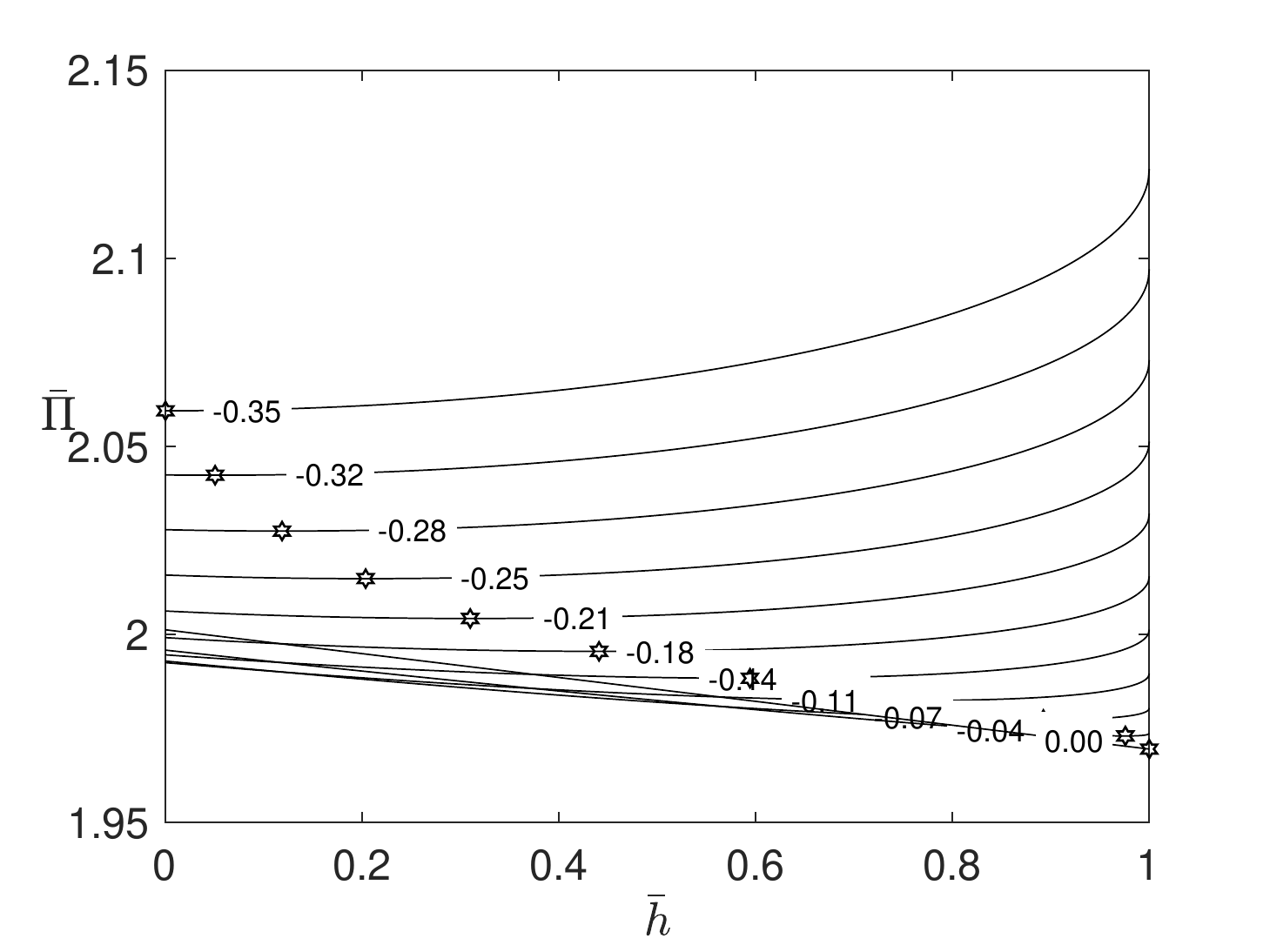}
        \caption{{\footnotesize Energy profiles  for varying $\bar{T}$ with stable equilibria at various temperatures.}}
        \label{sfig:SmoothEnergy}
    \end{subfigure}
    ~
    \begin{subfigure}[b]{0.48\textwidth}
        \includegraphics[width=\textwidth]{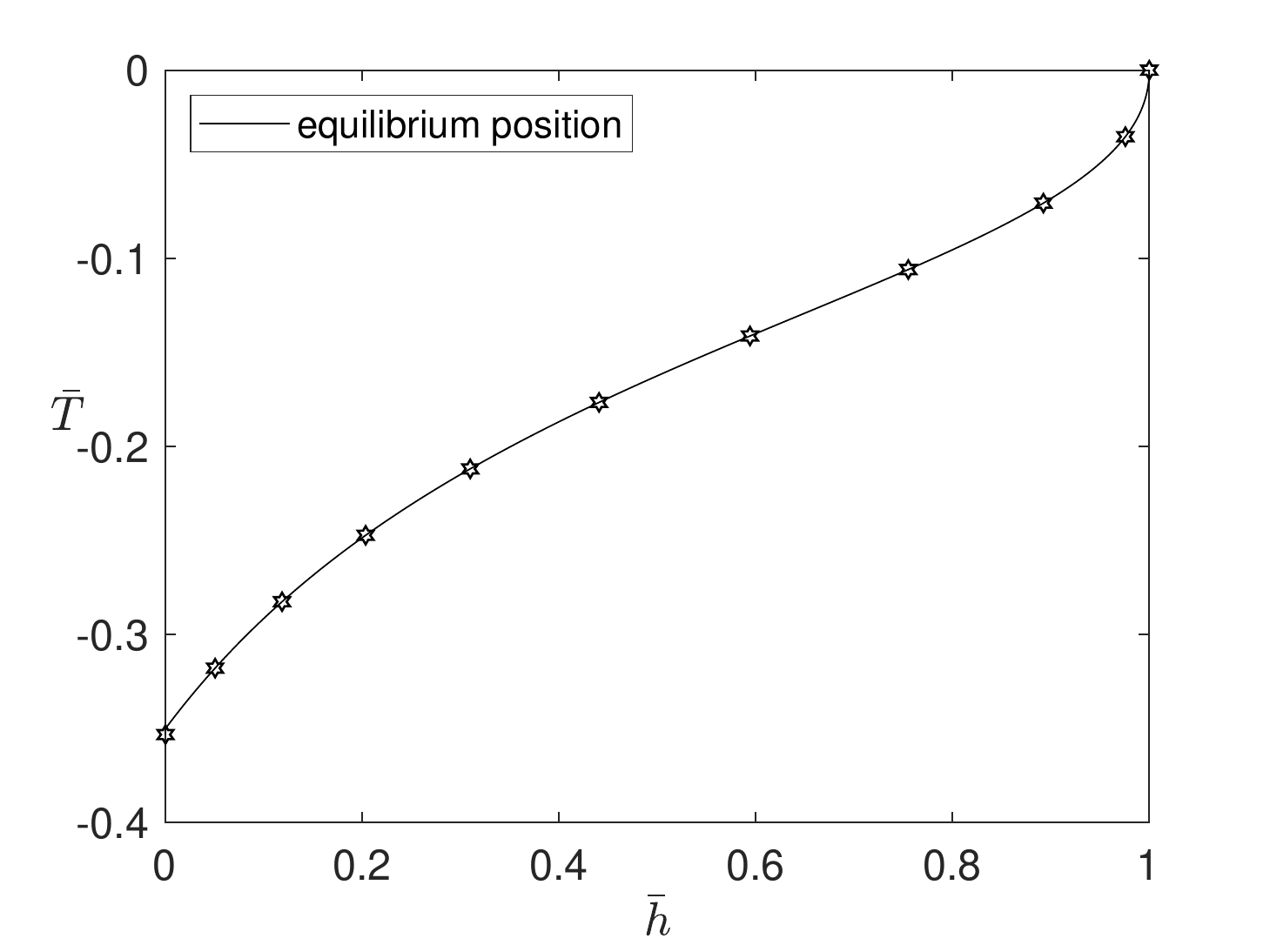}
        \caption{{\footnotesize Location of stable equilibrium during heating and cooling.}}
        \label{sfig:SmoothPos}
    \end{subfigure}   
    \caption{Smooth actuation from fully-coiled to fully-extended between critical temperatures $\T = -0.35$ and $\T = 0$.}
    \label{fig:Smooth}
\end{figure*}

\begin{figure*}
\centering
    \begin{subfigure}[b]{0.48\textwidth}
        \includegraphics[width=\textwidth]{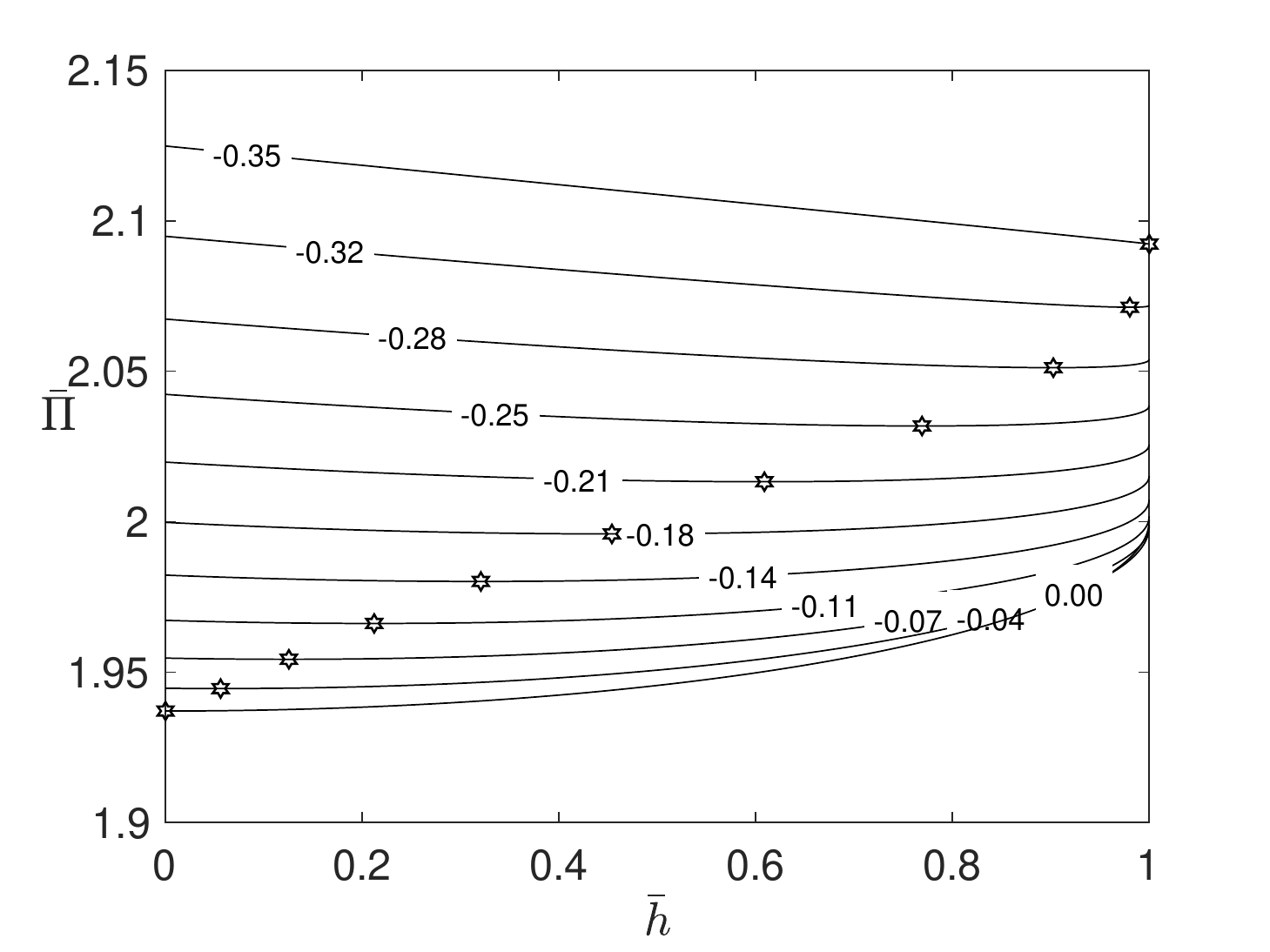}
        \caption{{\footnotesize Energy profiles  for varying $\bar{T}$ with stable equilibria at various temperatures.}}
        \label{sfig:SmoothREnergy}
    \end{subfigure}
    ~
    \begin{subfigure}[b]{0.48\textwidth}
        \includegraphics[width=\textwidth]{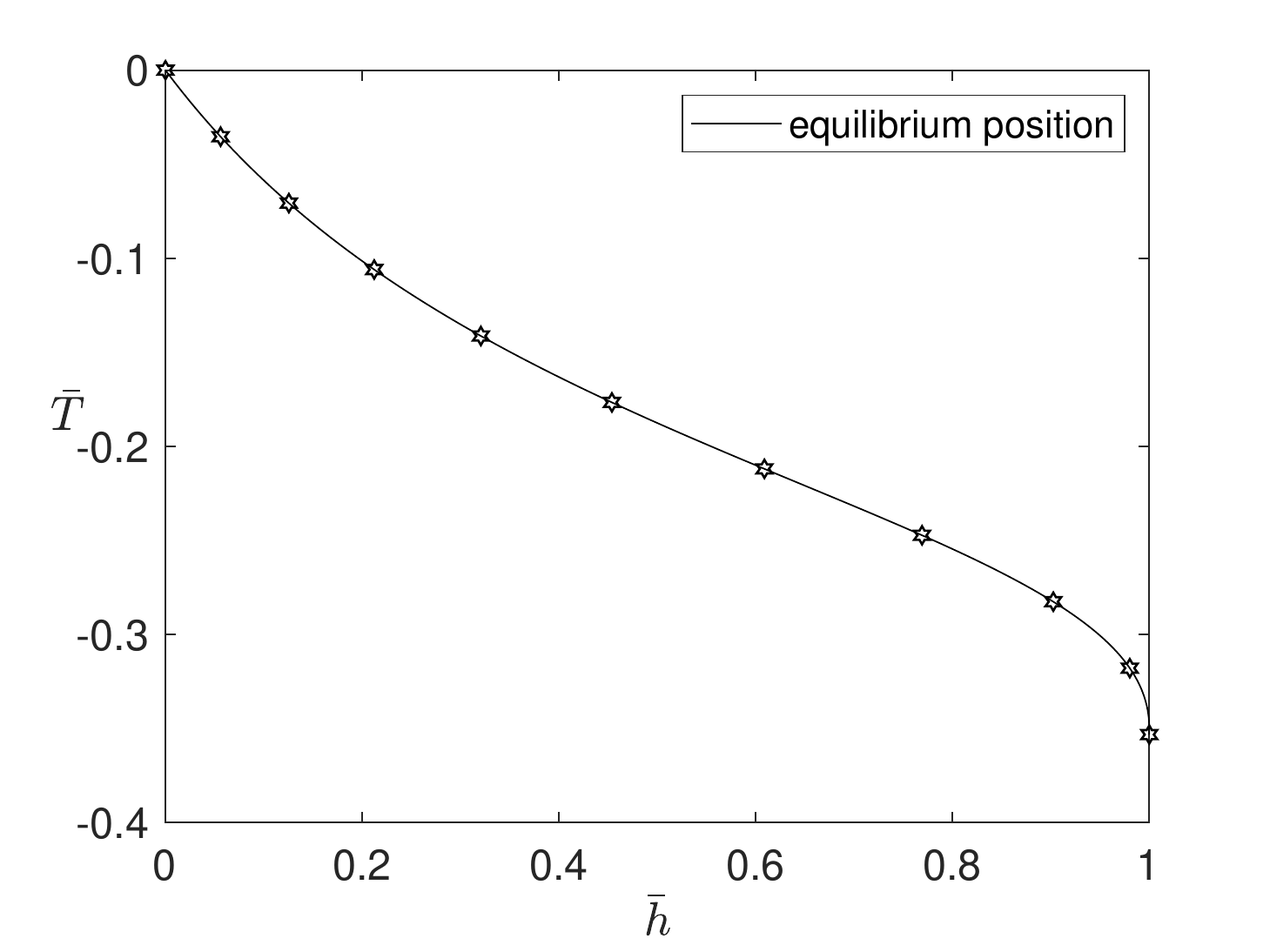}
        \caption{{\footnotesize Location of stable equilibrium during heating and cooling.}}
        \label{sfig:SmoothRPos}
    \end{subfigure}   
    \caption{Smooth actuation from fully-coiled to fully-extended between critical temperatures $\T = -0.35$ and $\T = 0$ illustrating negative coefficient of thermal expansion.}
    \label{fig:SmoothR}
\end{figure*}

To achieve snap-actuation at a critical temperature, the preceding examples exploit appearance or disappearance of equilibria at the extensional boundaries, i.e.\ at $\h \in \{0,1\}$. Alternatively, smooth actuation is obtained by exploiting stable inner equilibria, i.e.\ at $\h_0 \in (0,1)$. This behaviour is illustrated in Figures~\ref{fig:Smooth} and~\ref{fig:SmoothR}. (Existence of inner equilibria can be guaranteed by imposing suitable signs on the derivative of the energy on the boundary; see~\cite{Pirrera_et_al:2013} for details.)

In Figure~\ref{fig:Smooth}, the lattice expands smoothly under heating whereas in Figure~\ref{fig:SmoothR} it contracts smoothly under heating, i.e.\ has a negative coefficient of thermal expansion. Note that the CTEs demonstrated here, both positive and negative, are orders of magnitude larger in comparison to those of materials reported to have exceptionally large values \citep{Das2009,Miller2009,Fortes2011,Takenaka2012,Takenaka2017}; see~Figure~\ref{fig:Tailoring}.

\subsection{Thermal invariance}
\label{sec:thermal_invariance}

As noted in Section~\ref{sec:equilibria}, zero CTE at all temperatures is guaranteed by ${a}_{11} = {b}_{11} = 0$. More generally, the mechanics of the lattice is such that thermal invariance can be obtained within some ranges of temperatures. When the lattice is locked at $\h = 0$ or $\h = 1$, it is effectively thermally invariant until a critical temperature, either $\T_0$ or $\T_1$, is reached; see Figures~\ref{fig:Forward} and~\ref{fig:NoH}.

Because CTE, as defined in~\eqref{eq:CTE}, does not account for in-plane thermal expansion of the lattice's strips, the length $\h_0$ of the structure could change with temperature even for zero CTE. However, the possibility remains that the structure could be designed such that its CTE cancels the change in length due to the constituent materials' thermal expansion. Such a structure would be completely thermally invariant. 

On the other hand, when the magnitude of CTE is large, as in Section~\ref{sec:Smooth}, the change in length due to in plane expansion  
would typically be dwarfed by the large CTE.

\subsection{Dirac delta CTE}
\label{sec:Dirac_delta_CTE}

Further, we can devise a system which is mono-stable (with an interior equilibrium) but develops bi-stability at a critical temperature. To this end, consider the choice of parameters
\begin{align*}
{a}_{20}
 &= {b}_{20} = 0, \\
\frac{{a}_{10}}{{b}_{10}} &= \frac{{a}_{11}}{{b}_{11}} = \gamma \neq 0,
\end{align*} 
for some $\gamma \in \mathbb{R}$. 
Then, from~\eqref{eq:diff1a},
\begin{equation}
\frac{\partial\bar{\Pi}}{\partial \h}
 = \left( {a}_{10} + {a}_{11} \T \right) \left( 1 - \frac{\gamma\h}{\sqrt{1-\h^2}} \right),
\end{equation}
which is identically zero when $\T = -\frac{ {a}_{10} }{ {a}_{11} }$; otherwise the unique equilibrium is at $\h_0 = \frac{1}{\sqrt{1+\gamma^2}}$. The stability of this equilibrium can be determined from~\eqref{eq:diff2}:
\begin{equation}
\frac{\partial^2\bar{\Pi}}{\partial \h^2}
 = - \left( {a}_{10} + {a}_{11} \T \right)  \frac{1}{\left(1-\h^2\right)^{3/2}},
\end{equation}
from which it is clear that $\h_0$ is stable when $\T < -\frac{ {a}_{10} }{ {a}_{11} }$. At $\T = -\frac{ {a}_{10} }{ {a}_{11} }$ the energy becomes flat. If the temperature is increased further, the equilibrium at $\h_0$ becomes unstable and stable equilibria appear at the boundaries, i.e.\ $\h = 0$ and $\h = 1$. These boundary equilibria remain stable as the temperature is increased further. 

\begin{figure*}
\centering
    \begin{subfigure}[b]{0.48\textwidth}
        \includegraphics[width=\textwidth]{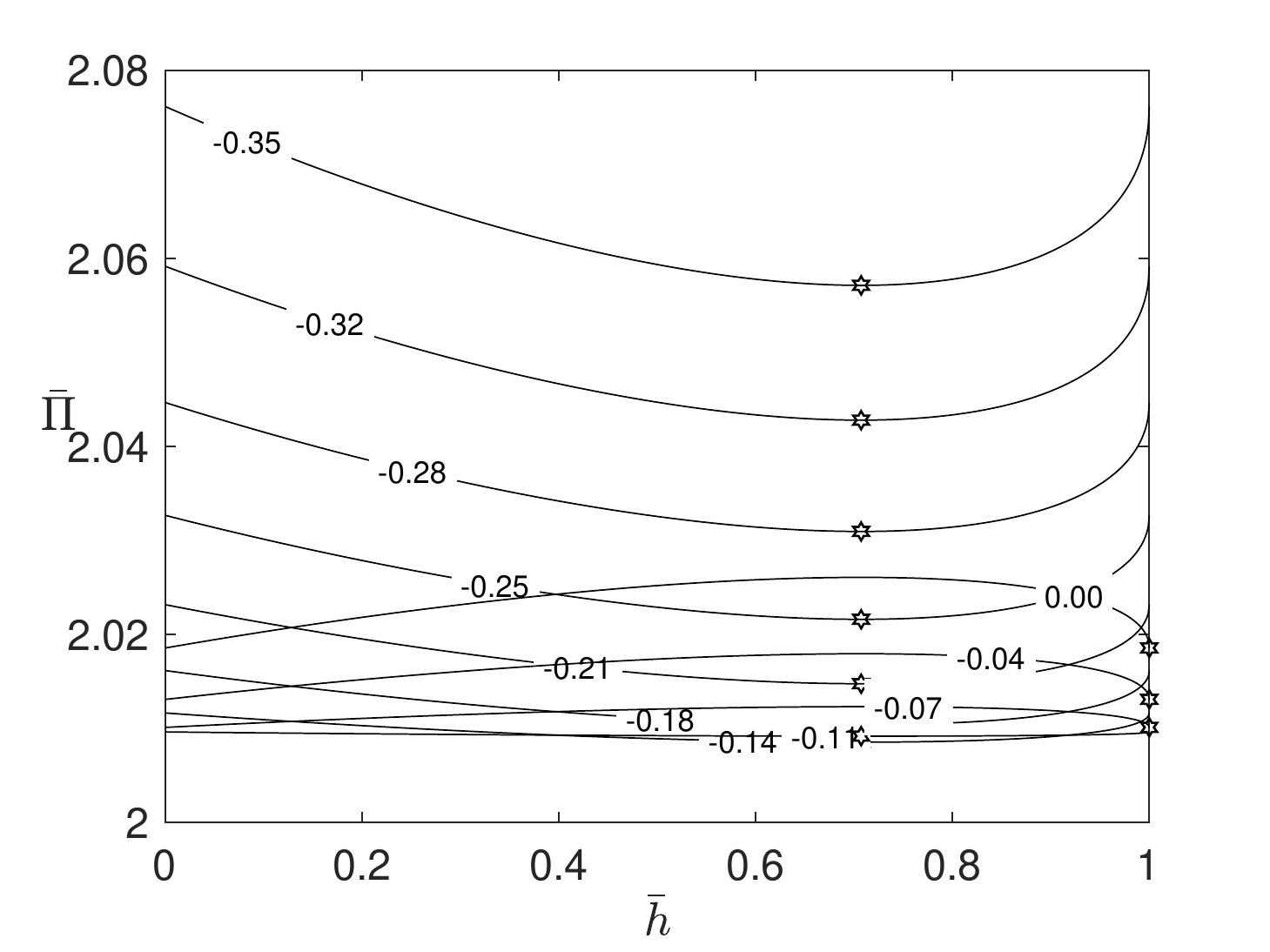}
         \caption{{\footnotesize Energy profiles  for varying $\bar{T}$ with stable equilibria at various temperatures.}}
        \label{sfig:CriticalE}
    \end{subfigure}
    ~
    \begin{subfigure}[b]{0.48\textwidth}
        \includegraphics[width=\textwidth]{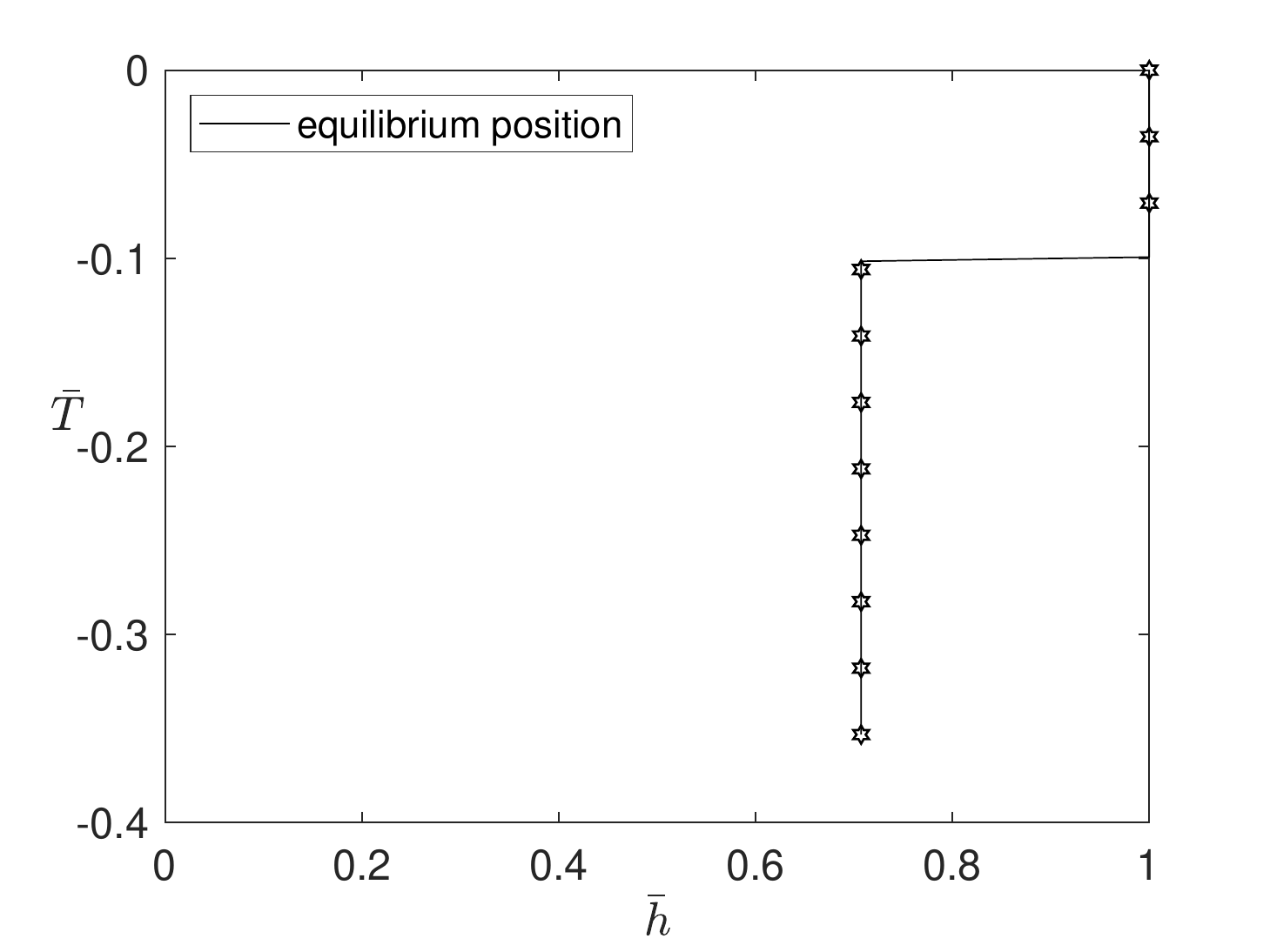}
        \caption{{\footnotesize Location of stable position during heating and cooling. }}
        \label{sfig:CriticalP}
    \end{subfigure}   
    \caption{Thermal invariance during heating of internal equilibrium position up to critical temperature $\T = -0.1$. }
    \label{fig:Critical}
\end{figure*}

\begin{figure*}
\centering
    \begin{subfigure}[t]{0.48\textwidth}
        \includegraphics[width=\textwidth]{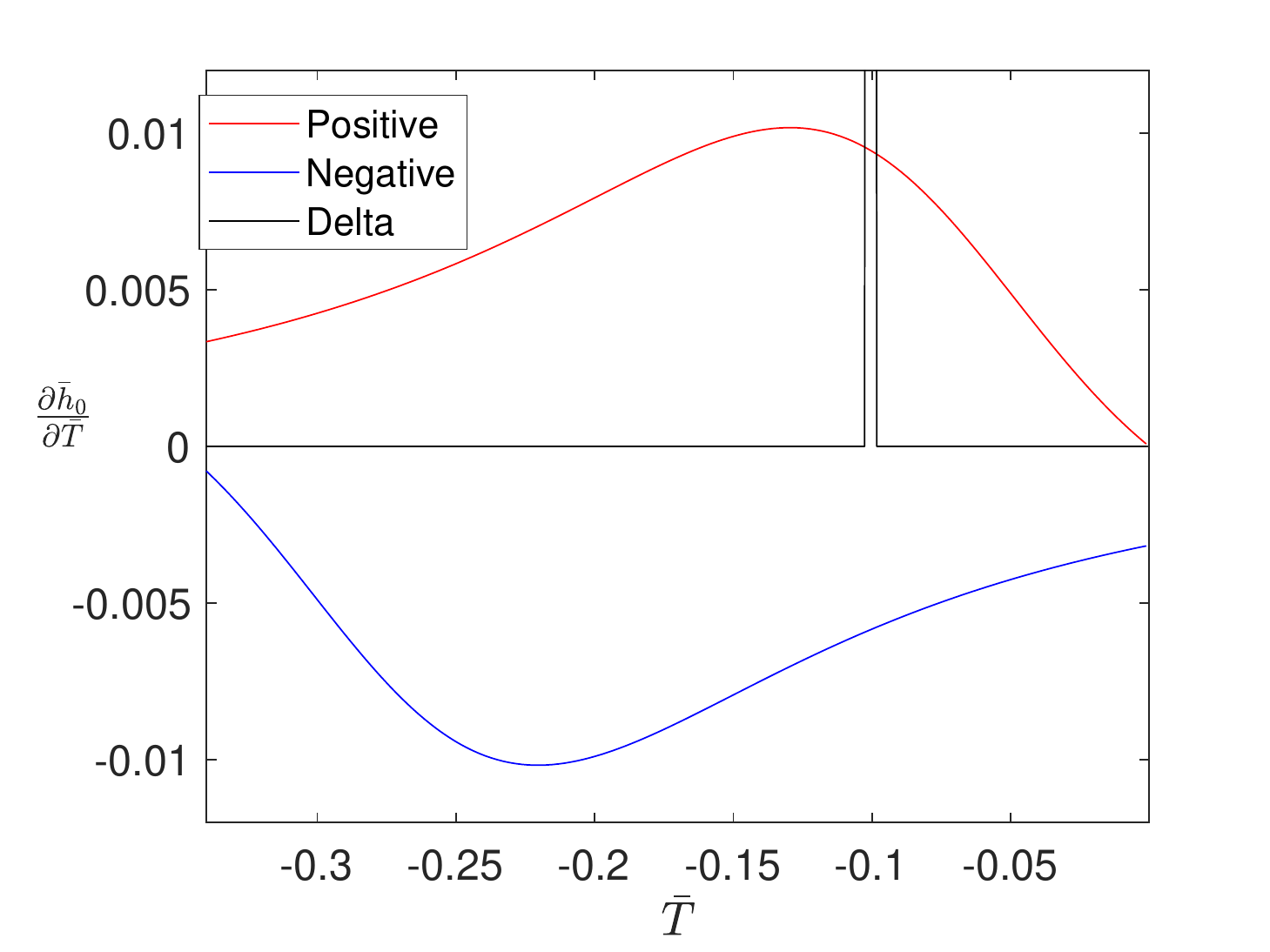}
        \caption{{\footnotesize Variation in the effective CTE.}}
        \label{sfig:CTEs}
    \end{subfigure}
    ~
    \begin{subfigure}[t]{0.48\textwidth}
        \includegraphics[width=\textwidth]{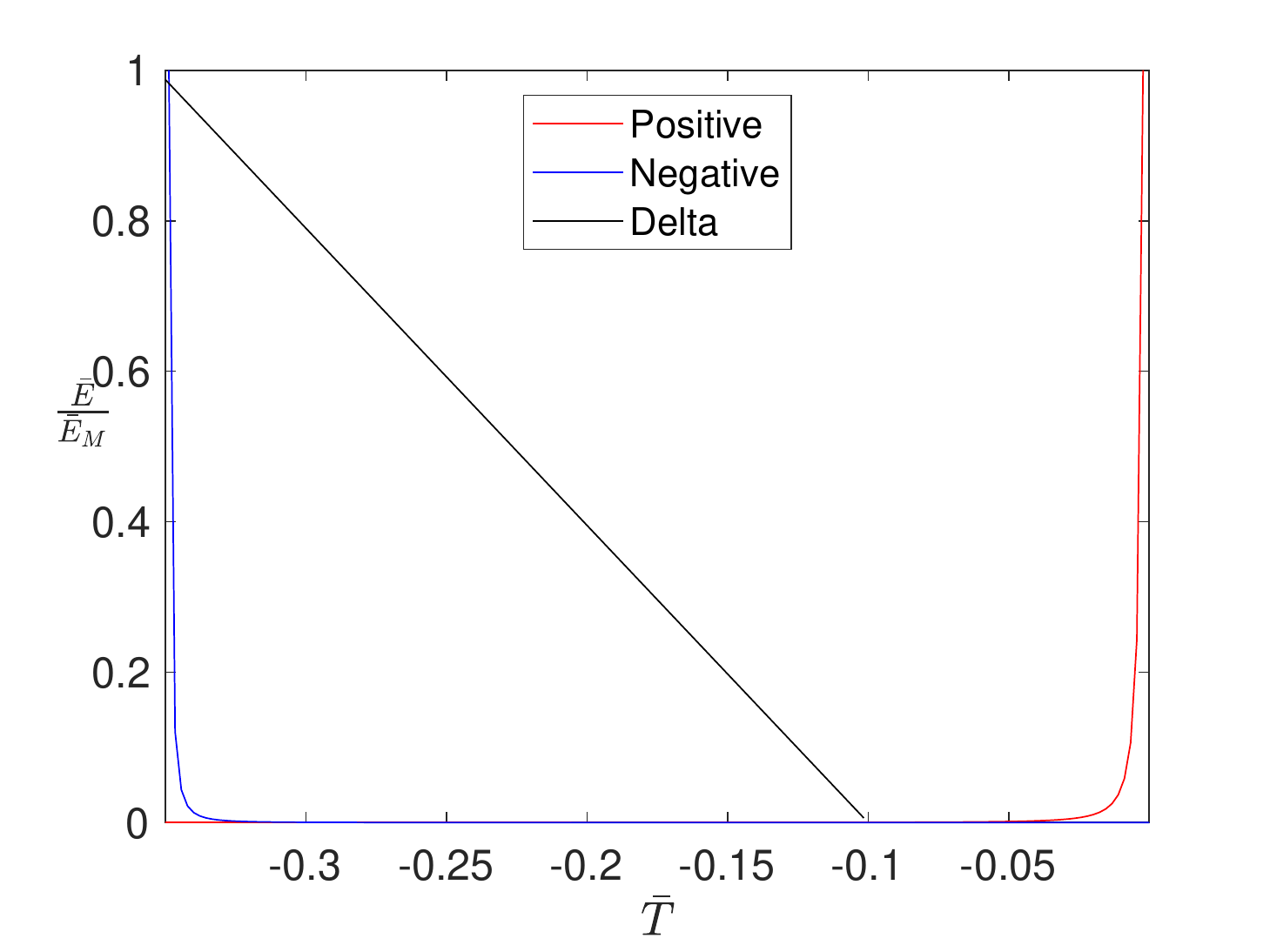}
        \caption{{\footnotesize Variation in effective initial modulus for varying $\T$ normalised by maximum observed value.}}
        \label{sfig:Modulus}
    \end{subfigure}   
    \caption{Effective CTE and stiffness of the lattice system for the three smooth-expansion cases shown in Figures~\ref{fig:Smooth}, \ref{fig:SmoothR}, and~\ref{fig:Critical}. Here the effective CTE is positive, negative and zero, respectively.}
    \label{fig:Tailoring}
\end{figure*}

As a consequence, the CTE is zero except at the critical temperature of $\T = -\frac{ {a}_{10} }{ {a}_{11} }$, see Figures~\ref{fig:Critical} and~\ref{fig:Tailoring} where $\gamma=1$. At this critical temperature, the system's length instantaneously changes from $\h_0 = \frac{1}{\sqrt{1+\gamma^2}}$ to either $0$ or $1$. Thus the CTE is a Dirac delta with a multiplicative pre-factor of either $-\frac{1}{\sqrt{1+\gamma^2}}$ or $(1-\frac{1}{\sqrt{1+\gamma^2}})$. Changing $\gamma$ allows the position $h_0$ to be tuned along with the multiplicative pre-factor.

Moreover, the flat energy landscape, with the consequent zero-stiffness, at critical temperature results in minimal actuation energy to move from one extension to another. This feature is desirable (quite apart from considerations of thermal expansion) in morphing applications since the system could be actuated in its zero-stiffness state, requiring minimal energy input, and then allowed to return to its normal state, with associated increase in stiffness.

\section{Proof-of-Concept Demonstrator}\label{sec:demonstraor}
\begin{figure*}
\centering
    \begin{subfigure}[b]{0.54\textwidth}
        \includegraphics[width=\textwidth]{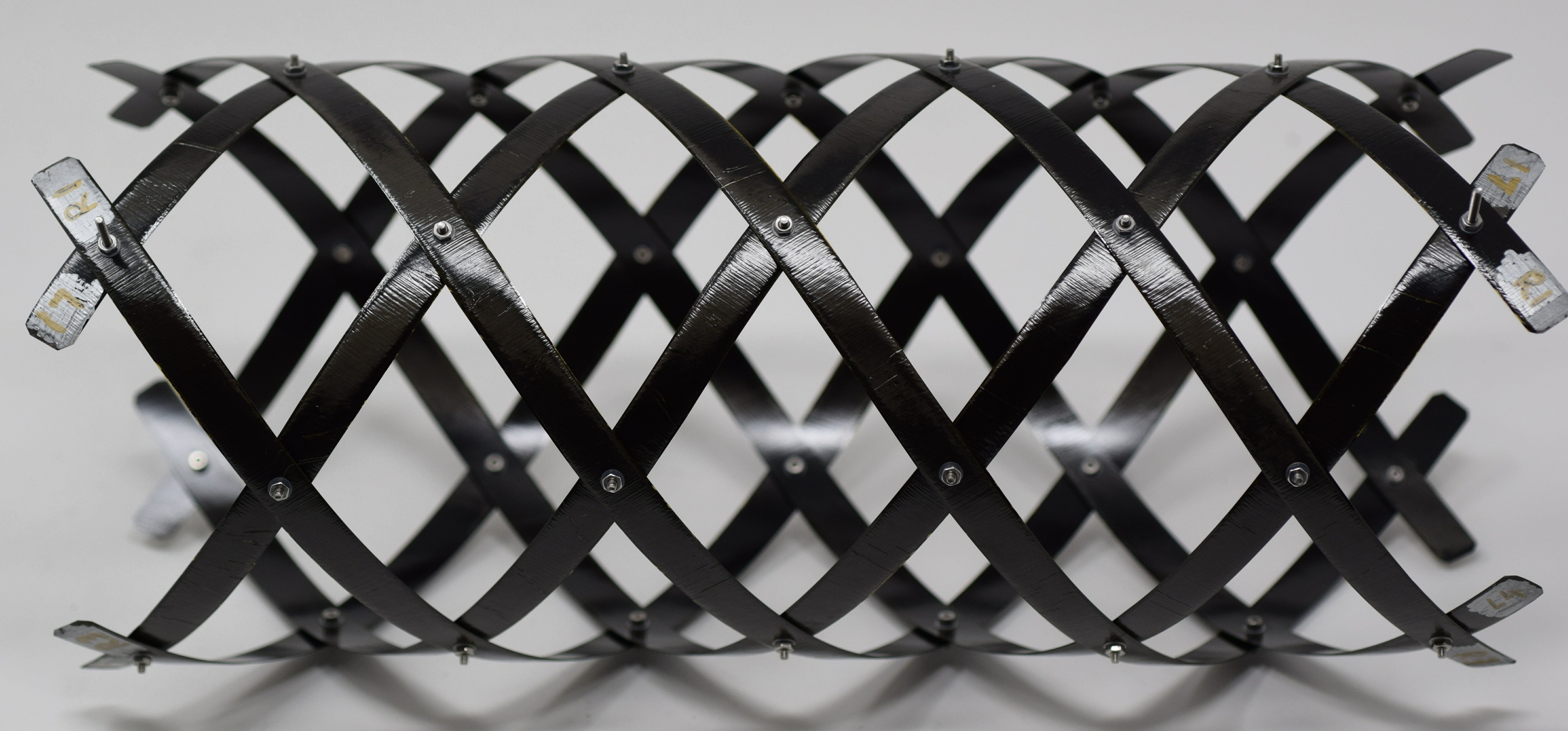}
        \caption{{\footnotesize Lattice at room temperature.}}
        \label{sfig:Dem1}
    \end{subfigure}
    ~
    \begin{subfigure}[b]{0.42\textwidth}
        \includegraphics[width=\textwidth]{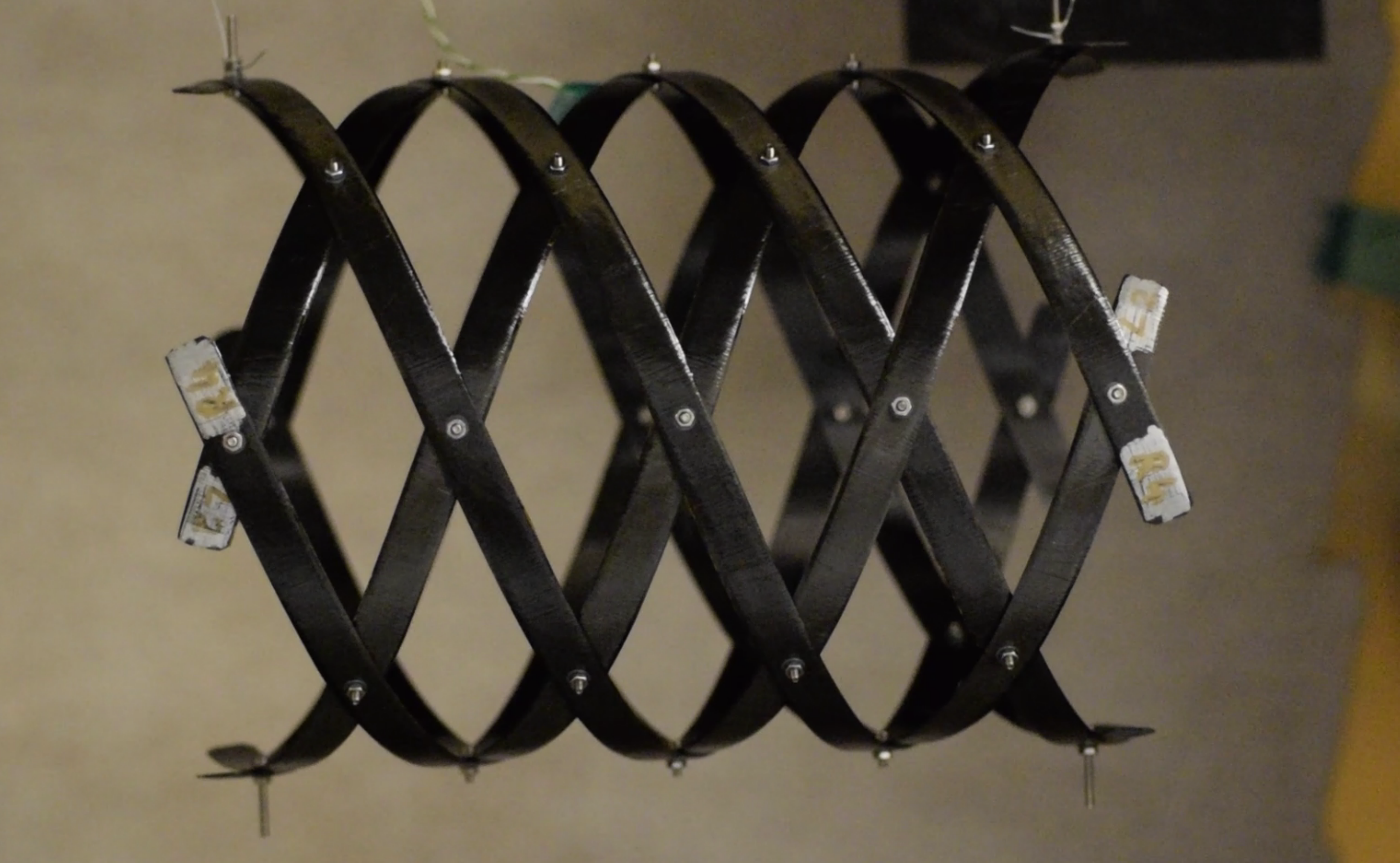}
        \caption{{\footnotesize Lattice in oven during experiment.}}
        \label{sfig:Dem2}
    \end{subfigure}   
    
    \begin{subfigure}[b]{0.80\textwidth}
        \includegraphics[width=\textwidth]{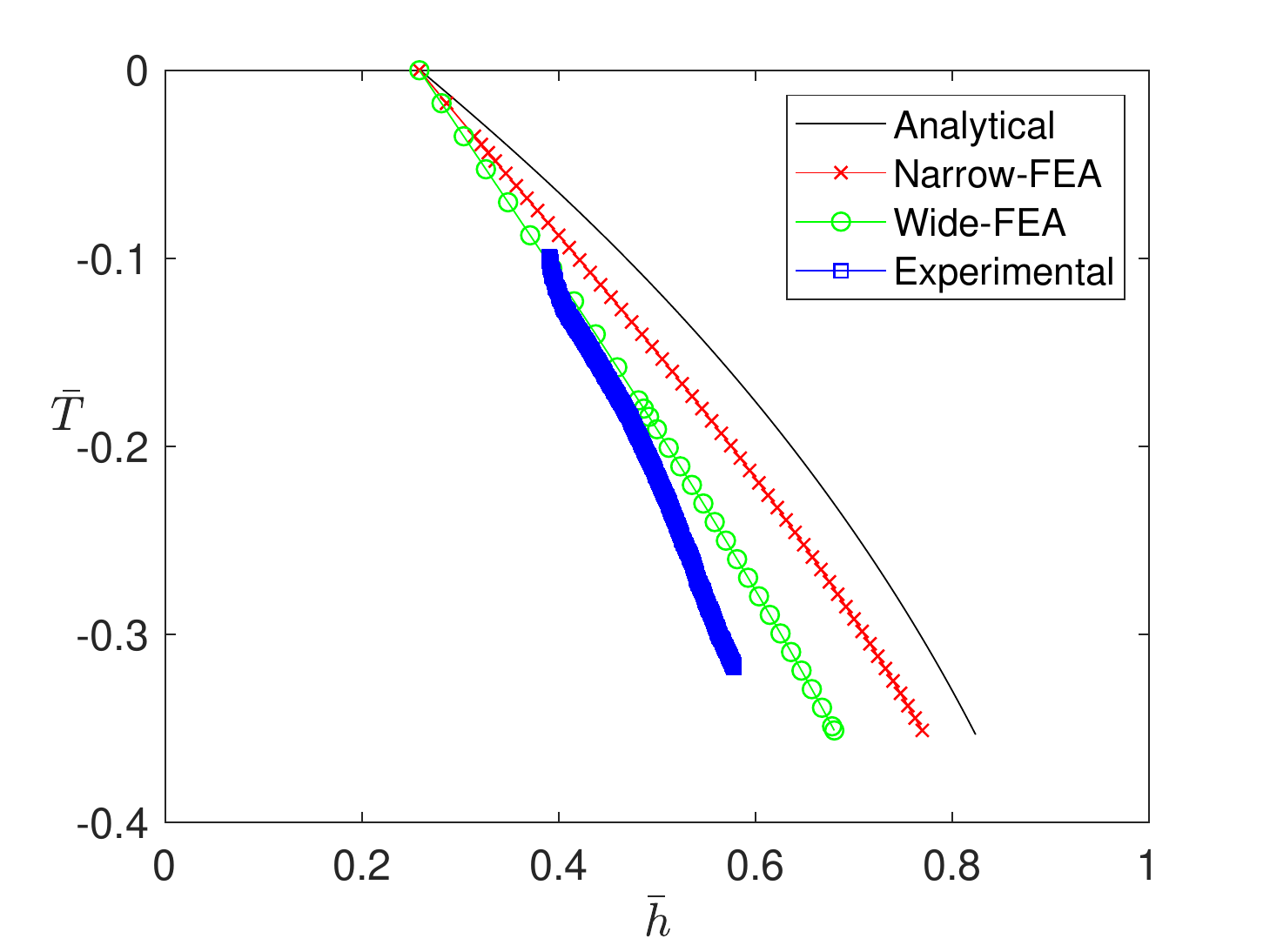}
        \caption{{\footnotesize Location of stable position during heating and cooling.}}
        \label{sfig:Dem3}
    \end{subfigure}
    ~
    \caption{Thermoelastic lattice demonstrator showing large contraction during heating.}
    \label{fig:Expansion}
\end{figure*}

In order to illustrate the feasibility of the proposed concept, we present a thermoelastic lattice demonstrator tuned to exhibit a large negative CTE, Figure \ref{sfig:Dem1}. The design specifications are selected for ease of manufacture rather than performance optimisation. The lattice is cured on a cylindrical tool and is constructed as described in Table~\ref{tab:latticeprops}. The strips are manufactured as tapes of carbon fibre reinforced plastic (IM7-8552), with the material properties obtained from literature~\citep{Ratcliffe2012} and in-house measurements. The thermal behaviour of the lattice was obtained by tracking its expansion from a heated state as it cooled whilst suspended horizontally in an oven; see Figure~\ref{sfig:Dem2}.

The width of the lattice's strips is $10$\,mm, which is larger than would be desirable in light of the one-dimensional strip behaviour assumed by the analytical model presented herein, but minimum sizes were dictated by current manufacturing capability. For this reason, we also include results obtained using an \textsc{abaqus} finite element (FE) analysis model (quasi-static, geometrically nonlinear analysis with \texttt{S4R} shell elements). Figure~\ref{sfig:Dem3} shows clear correlation between the FE, analytical and experimental results. As expected, the correlation between the analytical prediction and the FE model improves as the width of the strip is reduced from $10$\,mm (labelled `wide' in Figure~\ref{fig:Expansion}) to $0.5$\,mm, (labelled `narrow'). 

In real terms, the total lattice length contracts from $286$\,mm to $193$\,mm during a temperature increase from $309$\,K to $408$\,K. This response gives a representative CTE over the range of motion of $-3285\times10^{-6}\,\mathrm{K}^{-1}$. We remark that the effective CTE is over 100 times larger than $-30\times10^{-6}\,\mathrm{K}^{-1}$  often stated as a large negative CTE \citep{Takenaka2012}. Our result exceeds even typically extreme values reported in the literature $-515 \times10^{-6}\,\mathrm{K}^{-1}$ \citep{Das2009, Fortes2011}.


We note that this demonstrator does not represent the upper limit of what can be achieved using the thermoelastic lattice. Referring to the coefficients presented in table \ref{tab:properties}, and figure \ref{fig:SmoothR}, we are free to tune the transition temperatures $\T_0$ and $\T_1$, and therefore in principle achieve any average CTE for the temperature interval $(\T_1,\T_0)$.  Thus these experimental results demonstrate the potential of architectured thermoelastic materials to deliver extreme performance requirements inaccessible through traditional design approaches.

\begin{table}[]
    \centering
    \footnotesize 
    \begin{tabular}{rl|rl}
    \toprule
         $E_{11}$& 161 GPa  &
         $E_{22}$& 11.38 GPa \\
         $G_{12}$& 5.17 GPa &
         $\nu_{12}$ & 0.32 \\
         $\alpha_{11}$ & -0.1$\times10^{-6}\,\mathrm{K}^{-1}$ &
         $\alpha_{22}$ & 31.0$\times10^{-6}\,\mathrm{K}^{-1}$ \\
         $N$ & 4&
         $w_\pm$ & 10 mm \\
         Layup$_+$& [-15,-70,-35,90]&
         Layup$_-$& [+15,+70,+35,90]\\
         Ply Thickness & 0.1177 mm &
         Tool Radius & 76 mm\\
         Tool Angle & 15\degree&
         Cell Length & 62 mm \\
         \bottomrule
    \end{tabular}
    \caption{Properties of the thermoelastic lattice demonstrator}
    \label{tab:latticeprops}
\end{table}

\section{Conclusions}

Based on the model presented in Sections~\ref{sec:sec2} and~\ref{sec:sec3}, the analysis of Section~\ref{sec:responses} demonstrates how a structure, in this case a helical lattice, can be designed to mimic a one-dimensional thermoelastic material.

Section~\ref{sec:Forward} illustrates how the design parameters can be tuned to achieve effective behaviour which mimics that of phase-transforming materials such as SMAs. Indeed, we can more than mimic such materials; we can improve on their performance. For example, we can easily tune parameters like the transition temperatures, which are typically much more difficult to design at the crystalline micro-structure scale through a materials science approach. Further, as Section~\ref{sec:NoH} shows, it is no less easy to reduce and even eliminate hysteresis. In contrast, eliminating hysteresis in materials such as SMAs requires considerable expertise, cf.\ e.g.\ \citet{Cui2006}. With the proposed system, even the direction of actuation can be reversed, as noted in Section~\ref{sec:Smooth}. 

Turning to thermoelasticity more broadly, we can design for negative CTE and for magnitudes of CTE, both positive and negative, beyond what has been observed in nature; this is exemplified in Section~\ref{sec:Smooth}. In Section~\ref{sec:thermal_invariance}, we designed a system whose CTE is zero, as well as one whose CTE is a Dirac delta in Section~\ref{sec:Dirac_delta_CTE}. In Section~\ref{sec:demonstraor}, we illustrated the feasibility of constructing a thermoelastic lattice with a large negative CTE, experimentally demonstrating the potential of the proposed concept. By doing so, we highlighted the potential to obtain extreme response characteristics that are several orders of magnitude larger than typical existing materials.  

The examples presented are only a brief foray into the capacity to tune the coefficient of thermal expansion and extensional stiffness simultaneously. In particular, as Section~\ref{sec:thermal_invariance} demonstrates, the stiffness of a temperature-invariant equilibrium can itself be temperature-dependent. This holds out the promise of overcoming the difficulties incurred using traditional approaches to obtaining a system with desirable CTE and stiffness \citep{Jefferson_et_al:2009}.

In conclusion, we have outlined a structural route to attain and exceed capabilities which are normally associated with multi-phase materials such as SMAs or exotic structural materials with negative CTEs. In the language of \cite{Bhattacharya_James:2005}, we have a machine which acts as a material.

In this paper, we have exploited only a small portion of the design landscape that is opened up by the coupling between pre-stress, stiffness, thermal expansion and geometric nonlinearity. In particular, in future work, we aim to explore lattice systems that mimic multi-phase thin-films and three-dimensional solids. 

Finally, we note that, although the present approach focuses specifically on thermal actuation, any curvature-inducing field could be utilised instead; for example, electric fields in conjunction with piezoelectric materials.

\section*{}

\paragraph{Acknowledgements}
This research was supported by the Engineering and Physical Sciences Research Council (EPSRC) through the University of Bristol's Centre for Doctoral Training in Advanced Composites for Innovation and Science [Grant No. EP/L016028/1] and Alberto~ Pirrera is funded on an EPSRC Early Career Research Fellowship [Grant No. EP/M013170/1].

\paragraph{Data Access Statement}
The data necessary to support the conclusions are included in the paper.

\section*{References}
\bibliography{HelicesRefs}
\bibliographystyle{elsarticle-harv} 

\end{document}